\documentclass[superscriptaddress,nofootinbib,reprint,amsmath,amssymb,aps]{revtex4-2}
\usepackage{hyperref}
\usepackage{xcolor}
\usepackage{soul}
\usepackage{float}
\usepackage[normalem]{ulem}
\hypersetup{
    colorlinks,
    linkcolor={red},
    citecolor={blue!50!black},
    urlcolor={blue!80!black}
}
\usepackage[english=usenglishmax]{hyphsubst}
\usepackage{graphicx}
\usepackage{mathtools}
\usepackage{orcidlink}

\newcommand{\dd}{{\rm d}}

\def\apjl{ApJL}

\def\aap{A\&A }

\def\memras{MmRAS}

\begin{document}

\title[Fully metallic geodesic lenses as analog electromagnetic models of static and spherically symmetric gravitational fields]{Fully metallic geodesic lenses as analog electromagnetic models\\ of static and spherically symmetric gravitational fields}

\author{Enderson Falcón-Gómez\orcidlink{0000-0002-0008-0624}}%
\email{efalcon@pa.uc3m.es}
\affiliation{University Carlos III of Madrid. 28911 Leganés, Spain}
\author{Vittorio De Falco\orcidlink{0000-0002-4728-1650}}%
\email{v.defalco@ssmeridionale.it}
\affiliation{Scuola Superiore Meridionale, Largo San Marcellino, 80138 Napoli, Italy}
\affiliation{Istituto Nazionale di Fisica Nucleare, Sezione di Napoli, Via Cintia 80126 Napoli, Italy}
\author{Kerlos Atia Abdalmalak\orcidlink{0000-0002-9544-2412}}
\affiliation{University Carlos III of Madrid. 28911 Leganés, Spain}
\affiliation{Technical University of Madrid, 28040 Madrid, Spain}
\affiliation{Aswan University, 81542 Aswan, Egypt}
\author{Adrián Amor-Martín\orcidlink{0000-0002-6123-4324}}
\affiliation{University Carlos III of Madrid. 28911 Leganés, Spain}
\author{Valentín De La Rubia\orcidlink{0000-0002-2894-6813}}
\affiliation{University Carlos III of Madrid. 28911 Leganés, Spain}
\author{Gabriel Santamaría-Botello\orcidlink{0000-0003-4736-0030}}%
\affiliation{University of Colorado Boulder, Boulder, 80309 Colorado, USA }
\author{Luis Enrique García Muñoz\orcidlink{0000-0002-3619-7859}}%
\email{legarcia@ing.uc3m.es}
\affiliation{University Carlos III of Madrid. 28911 Leganés, Spain}

\date{\today}

\begin{abstract}
We demonstrate that a fully metallic and air-filled geodesic waveguide can be employed as an analog electromagnetic model of a static and spherically symmetric gravitational field. By following the Plebanski formalism, a space-time metric of the aforementioned type is firstly encoded into the electromagnetic properties of a flat space-time region in the form of an isotropic and radially varying refractive index distribution. Then, a three-dimensional, air-filled, and axially symmetric waveguide, composed of two equally spaced and curved metallic surfaces, is employed. Its shape is tailored such that the effective paths, followed by transverse electromagnetic beams of microwave radiation within this waveguide, result equivalent to null-geodesics taking place in the aforementioned refractive medium. This strategy avoids the need for a refractive medium, although it only allows to reproduce the space-time metric on the invariant plane. Two analog electromagnetic models of gravity, using the proposed approach, are designed to reproduce the metric of both a Schwarzschild black hole and a Morris-Thorne wormhole. The results from full-wave simulations demonstrate that a one-dimensional Gaussian beam faithfully follows a path completely equivalent to general relativistic null geodesics with a mean relative error within 4\%.
\end{abstract}
\maketitle

\section{Introduction}
\label{sec:intro}
General relativity (GR), introduced by Einstein in late 1915, is one of the most revolutionary theories of gravity capable of describing several gravitational phenomena at macroscopic scales. Over the years, GR has been successfully validated through different experiments, such as: tests in the weak field regime within the solar system \cite{Will1993,Will2014,dyson1919determination}; the first direct detection of the gravitational wave emission resulting from the merger of two black holes (BHs), announced by LIGO and Virgo collaborations in 2015 \cite{abbott2016observation}; the first image of the supermassive BH located at the center of the galaxy M87, achieved by the Event Horizon Telescope collaboration in 2019 \cite{akiyama2019first}.

Nevertheless, there exist astrophysical situations, whose scales remain outside our experimental capabilities. The strong gravitational field near the event horizon of a BH is one example. Although analytical methods (e.g., solving wave equations in curved space-times \cite{friedlander1975wave}) alongside numerical techniques (e.g., the ``3 + 1'' ADM formulation \cite{luis2001numerical}) can help to better understand gravitational phenomena in such cases, an interesting alternative to conduct astrophysical experiments consists of employing analog models of gravity.

In this vein, we focus on \emph{analog electromagnetic (EM) models of gravity}, which can be defined as a region $\Omega \in \mathbb{R}^3$ settled in flat space-time, whose EM properties, namely the dielectric permittivity $\overleftrightarrow \epsilon$, magnetic permeability $\overleftrightarrow \mu $, and magneto-electric coupling $\overleftrightarrow \gamma$ tensors provide the necessary and sufficient conditions for the resulting Maxwell equations in $\Omega$ to acquire the same mathematical structure as Maxwell equations in an empty curved space-time, characterized by a metric tensor $\overleftrightarrow g$. The importance of this equivalence relies on the fact that it can be used as a computational tool for mimicking the propagation of EM fields in curved space-times \cite{plebanski1960electromagnetic}, as well as it provides alternative ways to understand gravity and its interaction with EM fields. 

In this work, we analyse analog EM models of gravity to reproduce the space-time geometry of \emph{static and spherically symmetric gravitational fields}. This topic has experienced a rise in interest over the last 25 years. Among the works on this subject, worth mentioning are: the usage of hyperbolic metamaterials to mimic the metric of a Schwarzschild BH \cite{fernandez2016anisotropic}; homogenized composite materials to reproduce the Schwarzschild-(anti-)de Sitter metrics \cite{mackay2011realization}; isotropic and radially varying refractive index distributions to mimic celestial mechanic problems \cite{genov2009mimicking}, the paths of photons \cite{falcon2022propagation}, planets \cite{evans1996optical}, and the metric of the exterior and the interior Schwarzschild solutions \cite{xiao2023analogy}.

In particular, researchers exploit the last form of the analog EM model of gravity as a starting point to build up analog EM models of gravity based on geodesic lenses \cite{kunz1954propagation,rinehart1948solution,rinehart1952family}. Due to the spherical symmetry of the analog EM model based on the refractive medium, light paths will be contained in a plane \cite{falcon2022propagation}, which can be set coincident with the equatorial plane. In accordance with Fermat's principle \cite{kunz1954propagation}, a photon follows a path of least time influenced by the refractive medium. Instead, as the geodesic lens consists of a two-dimensional, non-euclidean, and axially-symmetric dielectric surface of finite thickness embedded in three-dimensions, it is the geometry that physically determines the effective electrical distance between neighboring points on the surface, and thus the complete path of a light ray. Therefore, appropriately shaping the surface of the geodesic lens, the path followed by light beams therein can be made equivalent to those taking place on the equatorial plane of the analog EM model based on the refractive medium and, consequently, also in the equatorial plane of the gravitational setting, where the latter model stems out. Among the studies on this topic, we highlight an analog optical model of space-time outside a Morris-Thorne wormhole (WH) \cite{he2020simulation} implemented by He and collaborators using a curved dielectric layer. Furthermore, there exist other works exploiting the geodesic lenses to explore wave propagation in curved ``space-times'' \cite{sheng2013trapping,schultheiss2010optics}. However, in those cases, the authors do not determine the shape of the geodesic lens from a specific space-time geometry, as outlined before, but rather, they rely on fabrication possibilities or specific geometric characteristics of the resulting geodesic waveguide's surface.

In this work, we resort to a fully-metallic geodesic lens \cite{myers1947parallel,rinehart1948solution,kunz1954propagation}, which comprises two axially symmetric curved and equally-spaced metallic surfaces, where the filling medium is air. The main advantage of this method is that the waveguide operates with the transverse electromagnetic mode (TEM) so that the wave impedance equals the intrinsic impedance of vacuum and EM wave propagation does not undergo dispersion, in contrast to a fully dielectric geodesic lens, which operates with a transverse magnetic mode \cite{pozar2011microwave}. Furthermore, this strategy also avoids the use of metamaterials or composites and leverages the capability of a geodesic lens to control EM wave propagation through the geometry. Apart from that, this approach affords three more advantageous aspects. Firstly, consistency conditions \cite{schuster2018bespoke} of space-time are naturally fulfilled as the filling medium is air and not a dielectric material. Secondly, these waveguides have been widely used for applications in the microwave frequency range ($3-300~\text{GHz}$) \cite{pozar2011microwave} such as: radar scanning \cite{myers1947parallel,rinehart1948solution} and antennas based on graded-index media \cite{johnson1962geodesic,kunz1954propagation,liao2018compact,fonseca2018water}. Lastly, microwave technology comprises a wide variety of techniques that can be used to implement these analog models of gravity.
 
To assess the feasibility of the geodesic lens approach, we design and simulate analog EM models corresponding to a Schwarzschild BH \cite{schwarzschild1916gravitationsfeld} and a Morris-Thorne WH \cite{morris1988wormholes}. In both scenarios, we feed the waveguide with a Gaussian beam \cite{goldsmith1998quasioptical} having a carrier in the microwave frequency range ($140~\text{GHz}$). The resulting field distributions within the waveguide prove that the beam faithfully follows the ray trajectories predicted by GR, confirming thus its trustable applicability for reproducing the metric on the invariant plane of static and spherically symmetric space-times.

This article is organized as follows: in Sec.~\ref{sec_the_analog_em_model_of_gravity} the theory of analog EM model of gravity is presented; in Sec.~\ref{sec:geodesic_lens_approach} the geodesic lens approach is introduced; the results of this work are presented and discussed in Sec. \ref{sec:results}; finally, conclusions and new perspectives are gathered in Sec. \ref{sec:conclusions}.
 
\emph{Notations and conventions}. The metric tensor is indicated by $g_{\mu\nu}$ having signature $(-,+,+,+)$, while $\eta_{\mu \nu} = {\rm diag}(-1,1,1,1)$ is the Minkowsky metric. Greek indices run as $0,1,2,3$, whereas Latin indices as $1,2,3$. Vectors, rank-2 tensors, and basis vectors are indicated as; $\vec{A}$; $\overleftrightarrow  A$, $A^{~\mu \nu}$, $A_{\mu \nu}$, or $A^{\mu}_\nu$; $\boldsymbol{\hat A}$, respectively. The matrices are indicated with $[A]$. We use the International System of Units. Finally, $\epsilon_0$ and $\mu_0$ are the vacuum permittivity and permeability, respectively. 

\section{The analog EM model of gravity}
\label{sec_the_analog_em_model_of_gravity}
In this section, we first define the static and spherically symmetric gravitational field (see Sec. \ref{sec:grav_field}), then the formalism behind the analog EM models of gravity is introduced (see \ref{sec:EM_analog}) and particularized for static and spherically symmetric gravitational fields (see Sec.~\ref{sec:analog:EM_model_static_spherically}
); finally, we consider a parallel-plate waveguide model that will be the starting point to introduce and construct the geodesic lens (see Sec. \ref{sec:isotropic_and_2d_analog_models_of_black_holes}). 

\subsection{Static and spherically symmetric gravitational fields in isotropic coordinates}
\label{sec:grav_field}
The interval of a generic static and spherically symmetric gravitational field in Cartesian isotropic coordinates\footnote{See pp. 175 in Ref. \cite{weinberg1972gravitation} for an introduction to the isotropic coordinates and their properties or Appendix \ref{sec:isotropic_coordinates} in this paper.} \cite{weinberg1972gravitation,fernandez2016anisotropic,falcon2022propagation,falcon2023interaction} ($ct,x,y,z$) is given by
\begin{equation} 
\dd s^2 = H(r)c^2\dd t^2+J(r) \left[ \dd x^2 +\dd y^2+\dd z^2\right],\label{eq:isotropic_metric_tensor}
\end{equation}

where $r^2 = x^2+y^2+z^2$, and $H(r)$ and $J(r)$ are the components of the metric tensor defining the space-time. For a Schwarzschild BH \cite{carroll2019spacetime,falcon2022propagation}, Eq. \eqref{eq:isotropic_metric_tensor} reads as\footnote{See Eq. (8.2.14) in Ref. \cite{weinberg1972gravitation}, for more details.}
\begin{equation}
\dd s^2 = -\frac{\left( 1-\frac{r_{\rm si}}{r}\right)^2}{\left( 1+\frac{r_{\rm si}}{r}\right)^2}c^2\dd t^2+\left( 1+\frac{r_{\rm si}}{r}\right)^4\left[ \dd x^2 +\dd y^2+\dd z^2\right],\label{eq:isotropic_metric_tensor_BH}
\end{equation}
for $r\geq r_{\rm si}$ with $r_{\rm si}$ being the Schwarzschild radius defining the event horizon \cite{gron2016celebrating}. For the Morris-Thorne WH \cite{morris1988wormholes,falcon2022propagation}, we have that the metric tensor reads as
\begin{equation}
\dd s^2 = -c^2\dd t^2+\left[ 1+\left(\frac{b_{\rm oi}}{r}\right)^2\right]^2\left[ \dd x^2 +\dd y^2+\dd z^2\right],\label{eq:isotropic_metric_tensor_WH}
\end{equation}
for $r\geq b_{\rm oi}$ with $b_{\rm oi}$ being the WH mouth size.

\subsection{The Einstein-Maxwell's equations and the analog EM model of gravity}
\label{sec:EM_analog}
The Einstein-Maxwell's equations model Electrodynamics in curved empty space-times \cite{landau2013classical}
\begin{subequations}
\begin{align}
&\frac{\partial F_{\mu \nu} }{\partial x^\gamma} + \frac{\partial F_{\gamma \mu} }{\partial x^\nu} + \frac{\partial F_{\nu \gamma} }{\partial x^\mu} = 0,
\label{eq:chmaxwellsequationsincurvedspacetimemaxwellequation01}\\
&\frac{\partial f^{\mu \nu} }{\partial x^\nu} = -\mu_0  \mathcal{J}^\mu_{\text{den}},
\label{eq:chmaxwellsequationsincurvedspacetimemaxwellequation02}
\end{align}
\end{subequations}
where $F_{\mu \nu} = F_{\mu \nu}(\mathcal{E}_{\rm i},\mathcal{B}_{\rm i})$ is the EM field-strength tensor, $f^{\mu \nu} = f^{\mu \nu}(\mathcal{D}^{\rm i},\mathcal{H}^{\rm i})$ is the excitation tensor, $\mathcal{J}^\mu_{\rm den} = \sqrt{-g} \mathcal{J}^\mu$ is the four-current vector density with $\mathcal{J}^{\mu} = \mathcal{J}^{\mu}(\rho,\mathcal{J}^{\rm i})  $ being the current four-vector, and $g = \det (g_{\mu \nu})$ with $g_{\mu \nu}$ being the metric tensor, see Appendix.~\ref{appendix_C}. The Cartesian coordinate system is used in the rest of the paper unless otherwise stated, i.e., $x^\mu$ runs over $t,x,y,z$. In addition, $\mathcal{E}_{\rm i}$, $\mathcal{B}_{\rm i}$, $\mathcal{D}^{\rm i}$, $\mathcal{H}^{\rm i}$, $\mathcal{J}^{\rm i}$ , and $\rho$ are the covariant components of the three-vectors electric field intensity and magnetic flux, the contravariant components of the three-vectors electric flux density, magnetic field intensity and current density, and lastly, the charge density, respectively. Here,  $\rm i$ runs over $x,y,z$.
 
In 1960, Plebanski formulated a non-covariant version of the above Maxwell equations \cite{plebanski1960electromagnetic}. The mathematical structure of these equations resembles Maxwell's equations, modeling the propagation of EM fields in matter media framed in flat space-times, as will be better discussed later. The main contribution of Plebanski \cite{plebanski1960electromagnetic} is to have derived the set of space-time constitutive relations relating the components of field strength and excitation tensors. In matrix notation, they read as
\begin{subequations} \label{eq:constitutive}
\begin{align}
[{\mathcal{D}}] &= [\tilde{\epsilon}][{\mathcal{E}}]+[\tilde{\gamma}] [{\mathcal{H}}],\label{eq:subsection:constitutive_relationship01}\\
[{\mathcal{B}}] &= [\tilde{\mu}][{\mathcal{H}}]-[\tilde{\gamma}][{\mathcal{E}}],\label{eq:subsection:constitutive_relationship02}
\end{align}
\end{subequations}
where the constitutive matrices\footnote{In this work, only Cartesian coordinates are contemplated. Although the Plebanski formalism can be extended to other curvilinear and orthogonal coordinate systems, we refrain from doing so for the sake of simplicity.} $[\tilde{\epsilon}]$, $[\tilde{\mu}]$, and $[\tilde{\gamma}]$  are \cite{plebanski1960electromagnetic,landau2013classical,schuster2017effective}
\begin{subequations} \label{eq:consistency}
\begin{align}
\frac{[\tilde{\epsilon}]}{\epsilon _0} = \frac{[\tilde{\mu}]}{\mu _0} &= -\frac{\sqrt{-g}}{g_{00}} \begin{pmatrix}
g^{\rm xx} & g^{\rm xy} & g^{\rm xz} \\
g^{\rm yx} & g^{\rm yy} & g^{\rm yz} \\
g^{\rm zx} & g^{\rm zy} & g^{\rm zz} 
\end{pmatrix},
\label{eq:subsection:Constitutive_relationships_for_non-covariant_Maxwell_equations_in_curved_spacetime_C}\\
[\tilde{\gamma}] &= \frac{1}{c g_{\rm tt}} \begin{pmatrix}
    0   & -g_{\rm tz} &  g_{\rm ty}  \\
 g_{\rm tz} &    0    & -g_{\rm tx}  \\
-g_{\rm ty} & g_{\rm tx} & 0  \\
\end{pmatrix} .
 \label{eq:summary_gamma_analog_EM_model}\end{align}
 \label{eq:constitutive_matrices}
\end{subequations}
Altogether, the non-covariant Maxwell's equation and, in particular, Eqs.~\eqref{eq:subsection:constitutive_relationship01} and \eqref{eq:subsection:constitutive_relationship02} set up the basis for the \emph{analog EM model of gravity}. Their physical interpretation is that a gravitational field affects the propagation of EM fields in the same way as a reciprocal bi-anisotropic material \cite{odell1962electrodynamics} settled in flat space-time does, while the constitutive matrices $[\tilde{\epsilon}]$, $[\tilde{\mu}]$, and $[\tilde{\gamma}]$ play the role of effective dielectric permittivity, magnetic permeability, and magneto-electric coupling tensors\footnote{We highlight that the constitutive matrices are not tensors \cite{schuster2017effective}.}, respectively. 

In view of what was stated, we formulate the Plebanski formalism as follows: let us consider the ``\emph{gravitational domain}'' consisting of a manifold $\mathcal{M}$ endowed with a metric $g_{\mu \nu}$, coordinates $x^{\mu}$, and an EM configuration $\left(F_{\mu \nu}, f^{\mu \nu},\mathcal{J}^{\mu}_{\rm den}, \chi^{\alpha \beta \mu \nu} \right)$. The tensor $ \chi^{\alpha \beta \mu \nu} $ is the constitutive tensor \cite{post1962formal}, containing the space-time constitutive relationships between $F_{\mu \nu}$ and $f^{\mu \nu}$. It can be derived by representing Eqs.~\eqref{eq:constitutive_matrices} in tensorial form, that is
\begin{equation}
f^{\alpha \beta} = \frac{1}{2} \chi^{\alpha \beta \mu \nu} F_{\mu \nu}.
\end{equation}
For further details on the constitutive tensor and its properties, the reader can consult Refs.~\cite{post1962formal,fathi2016century,thompson2011completely}. Next, let us consider the ``analog domain'' consisting of another manifold $\mathcal{\hat{M}}$, which is endowed with a Minkowski metric $\eta_{\mu \nu}$, coordinates $\hat{x}^{\mu}$, and an equivalent EM configuration $(\hat{F}_{\mu \nu}, \hat{f}^{\mu \nu},\mathcal{\hat{J}}^{\mu}_{\rm den},\hat{\chi}^{\alpha \beta \mu \nu})$. Note that the symbols used here have the same meaning as stated for the gravitational domain, while the hats indicate they are framed in the analog domain. Moreover, due to the absence of gravity in the analog domain, the constitutive tensor encodes constitutive relationships of the matter medium. The EM properties and constitutive relationships of such a medium are assumed for the moment unknown. 
 
The premise of Plebanski's formalism is that both the gravitational and analog domains must be electrodynamically equivalent. To this end, a one-to-one map $\mathcal{T}:~\mathcal{\hat{M}} \to {\mathcal{M}}$ is defined\footnote{We use the pull-back formalism from Thompson et al. \cite{fathi2016century,thompson2011completely} to define the map between the gravitational and analog domains.}, with

\begin{equation}
\mathcal{T}\left(c\hat{t},\hat{x},\hat{y},\hat{z}\right) =  \left(ct,x,y,z \right) = \left(c\hat{t},\hat{x},\hat{y},\hat{z} \right),\label{eq:transformation_T}
\end{equation}
As an example of the operation principle of this map: $F_{\alpha \beta} \Big\rvert_{{x}^{\alpha}}  = \mathcal{T}(\hat{F}_{\mu \nu}\Big\rvert_{\hat{x}^{\eta}})$ and  $\mathcal{T}(\hat{F}_{\mu \nu}) = \frac{\partial {x}^{\alpha}}{\partial \hat{x}^\mu} \frac{\partial {x}^{\beta}}{\partial \hat{x}^\nu}\hat{F}_{\mu \nu}$ is an appropriate tensor transformation relationship for the argument. Its pull-back \mbox{$\mathcal{T}^*:~\mathcal{M}\to {\mathcal{\hat{M}}}$} reads as

\begin{equation}
\mathcal{T}^*\left(ct,x,y,z \right) = \left(c\hat{t},\hat{x},\hat{y},\hat{z}\right) = \left(ct,x,y,z \right),\label{eq:transformation_T*}
\end{equation}
which operates in the opposite sense  \cite{fathi2016century,thompson2011completely}. As a result, we obtain the following correspondences:
\begin{subequations}
\begin{align}
\hat{F}_{\alpha \beta} \Big\rvert_{\hat{x}^{\eta}} &=  {F}_{\alpha \beta} \Big\rvert_{\mathcal{T}(\hat{x}^{\eta})}     , \label{eq:transformation_rule_Fstrength_munu_B}\\
\hat{f}^{\alpha \beta} \Big\rvert_{\hat{x}^{\eta}} &= {f}^{\alpha \beta} \Big\rvert_{\mathcal{T}(\hat{x}^{\eta})}     ,\label{eq:transformation_rule_fexcitation_munu_B}\\
\hat{J}^{\alpha}_{\rm den} \Big\rvert_{\hat{x}^{\eta}} &=  {J}^{\alpha}_{\rm den}  \Big\rvert_{\mathcal{T}(\hat{x}^{\eta})}     .\label{eq:transformation_rule_current_munu_B}\\
\hat{\chi}^{\alpha \beta \mu \nu} \Big\rvert_{\hat{x}^{\eta}}& = {\chi}^{\alpha \beta \mu \nu} \Big\rvert_{\mathcal{T}(\hat{x}^{\eta})}     .\label{eq:transformation_rule_chi_munu_B}
\end{align}
\label{eq_transformation_rule_pull_back_B}
\end{subequations}
See the details of the above derivations in Appendix.~\ref{appendix_C}. The key result here is the transformation relationship \eqref{eq:transformation_rule_chi_munu_B}, from which two conclusions may be derived. First, the constitutive relationships of the material medium in which EM fields propagate in the analog domain are of the same type of Eqs.~\eqref{eq:subsection:constitutive_relationship01} and \eqref{eq:subsection:constitutive_relationship02}, that is
\begin{subequations} \label{eq:constitutive_2}
\begin{align}
[{\mathcal{\hat{D}}}] &= [\hat{\epsilon}][{\mathcal{\hat{E}}}]+[\hat{\gamma}] [{\mathcal{\hat{H}}}],\label{eq:subsection:constitutive_relationship01_C}\\
[{\mathcal{\hat{B}}}] &= [\hat{\mu}][{\mathcal{\hat{H}}}]-[\hat{\gamma}][{\mathcal{\hat{E}}}],\label{eq:subsection:constitutive_relationship02_C}
\end{align}
\end{subequations} 
where $[\hat{\epsilon}]$ is the dielectric permittivity tensor, $[\hat{\mu}]$ is the magnetic permeability tensor, and $[\hat{\gamma}]$ is the magneto-electric coupling tensor of the material medium in the analog domain. Secondly, the constitutive matrices in the gravitational domain are mapped into the constitutive tensors in the analog domain as follows 
 \begin{subequations} \label{eq:constitutive}
\begin{align}
[\hat{\epsilon}]^{\rm i j} \Big\rvert_{\mathcal{T}(\hat{x}^{\eta})} &= [\tilde{\epsilon}]^{\rm i j} \Big\rvert_{\hat{x}^{\eta}}, \label{eq:subsection:constitutive_relationship_eps_B}\\
[\hat{\mu}]^{\rm i j} \Big\rvert_{\mathcal{T}(\hat{x}^{\eta})} &= [\tilde{\mu}]^{\rm i j} \Big\rvert_{\hat{x}^{\eta}},\label{eq:subsection:constitutive_relationship_mu_B}\\
[\hat{\gamma}]^{\rm i j} \Big\rvert_{\mathcal{T}(\hat{x}^{\eta})}&= [\tilde{\gamma}]^{\rm i j} \Big\rvert_{\hat{x}^{\eta}},\label{eq:subsection:constitutive_relationship_gamma_B}
\end{align}
\end{subequations}
by substituting Eqs.~\eqref{eq_transformation_rule_pull_back_B} into the Maxwell's equations of the analog domain using the three-vector notation
\begin{subequations}
\begin{align}
\nabla \cdot \vec{\mathcal{\hat{D}}}&= \boldsymbol{\rho}_{\rm den},
\label{eq:maxwell_A_analog}\\
\nabla \cdot \vec{\mathcal{\hat{B}}} &= 0,
\label{eq:maxwell_B_analog}\\
\nabla \times \vec{\mathcal{\hat{H}}} &= \vec{\mathcal{\hat{J}}}_{\rm den}+\frac{\partial  \vec{\mathcal{\hat{D}}} }{\partial {\hat{t}}},
\label{eq:maxwell_C_analog}\\
\nabla \times \vec{\mathcal{\hat{E}}} &= -\frac{\partial \vec{\mathcal{\hat{B}}}}{\partial {\hat{t}}}.
\label{eq:maxwell_D_analog}
\end{align}\label{eq:maxwell_equations_flat_space-time}
\end{subequations}
The ensuing set of differential equations will assume the same structure as the non-covariant Maxwell's equations obtained by Plebanski \cite{plebanski1960electromagnetic} for EM fields propagating in the gravitational domain. In this way, the formal equivalence between EM wave propagation in empty curved space-time and matter media in flat space-time is established.

As final remarks, a discussion on the covariance properties of the Plebanski formalism and further possibilities of the pull-back transformation is beyond the scope of this paper. The reader can consult Refs. \cite{fathi2016century,thompson2010dielectric}.
\subsection{The analog EM model of static and spherically symmetric gravitational fields}\label{sec:analog:EM_model_static_spherically}

Having considered the Plebanski formalism, the analog EM model of gravity of static and spherically symmetric space-times in isotropic coordinates, cf. Eq.~\eqref{eq:isotropic_metric_tensor}, is

\begin{subequations}
\begin{align}
\frac{[\epsilon]^{\rm ij}}{\epsilon _0}& = \frac{[\mu]^{\rm ij}}{\mu _0} = n \left ( r \right ) \delta ^{\rm ij},
\label{eq:consistency_condition_A}\\
[\gamma]^{\rm ij} &= [0],\label{eq:consistency_condition_B}
\end{align}\label{eq:consistency_conditions}
\end{subequations}
where the isotropic refractive index distribution
\begin{equation}
n  \left( r \right)  = \sqrt{ \dfrac{-J(r)}{H(r)}},\label{eq:refractive_index_general}
\end{equation}
monotonically decreases with the radial distance being unity at infinity, and $\delta ^{\rm ij}$ is the Kronecker delta \cite{carroll2019spacetime}. Thus, the analog EM model of the gravitational field of a Schwarzschild BH (see Eq.~\eqref{eq:isotropic_metric_tensor_BH}) and a Morris-Thorne WH (see Eq.~\eqref{eq:isotropic_metric_tensor_WH}) consist of the following isotropic refractive index distributions and consistency conditions \eqref{eq:consistency_conditions}
\begin{subequations}
\begin{align}
{n} (r) &= \frac{\left( 1+\frac{r_{\rm si}}{ {r}}\right)^3}{1-\frac{r_{\rm si}}{ {r}}}\hspace{2cm}{\rm for}~{r}\geq r_{\rm si},\label{eq:refraction_index_schwarzschild}\\
n(r) &= 1+ \left(\frac{b_{\rm oi}}{ r}\right)^2\hspace{1.75cm}\text{for}~r \ge b_{\rm oi}.
\label{eq:ellisRefractionIndex}
\end{align}
\end{subequations} 

It is worth noticing that the refractive index distribution shown in Eq.~\eqref{eq:refraction_index_schwarzschild} diverges on the surface $\rho = r_{\rm si}$. Such behavior, however, is physically correct and reflects the fact that this surface is approached by EM radiation asymptotically, being thus an analog event horizon. On the other hand, for the Morris-Thorne WH, see Eq.~\eqref{eq:ellisRefractionIndex}, the analog WH mouth is defined by the surface $\rho = b_{\rm oi}$. Here, we conceive the analog EM model of a Morris-Thorne WH as two separated universes, whose refractive index distributions are given by Eq.~\eqref{eq:ellisRefractionIndex}. These universes are electromagnetically connected through the WH mouth by imposing the continuity of the tangential components of the EM field over $\rho = b_{\rm oi}$.
 
\subsection{Parallel-plate waveguide analog EM model}
\label{sec:isotropic_and_2d_analog_models_of_black_holes}

\begin{figure*}
\centering
\includegraphics[width=12 cm]{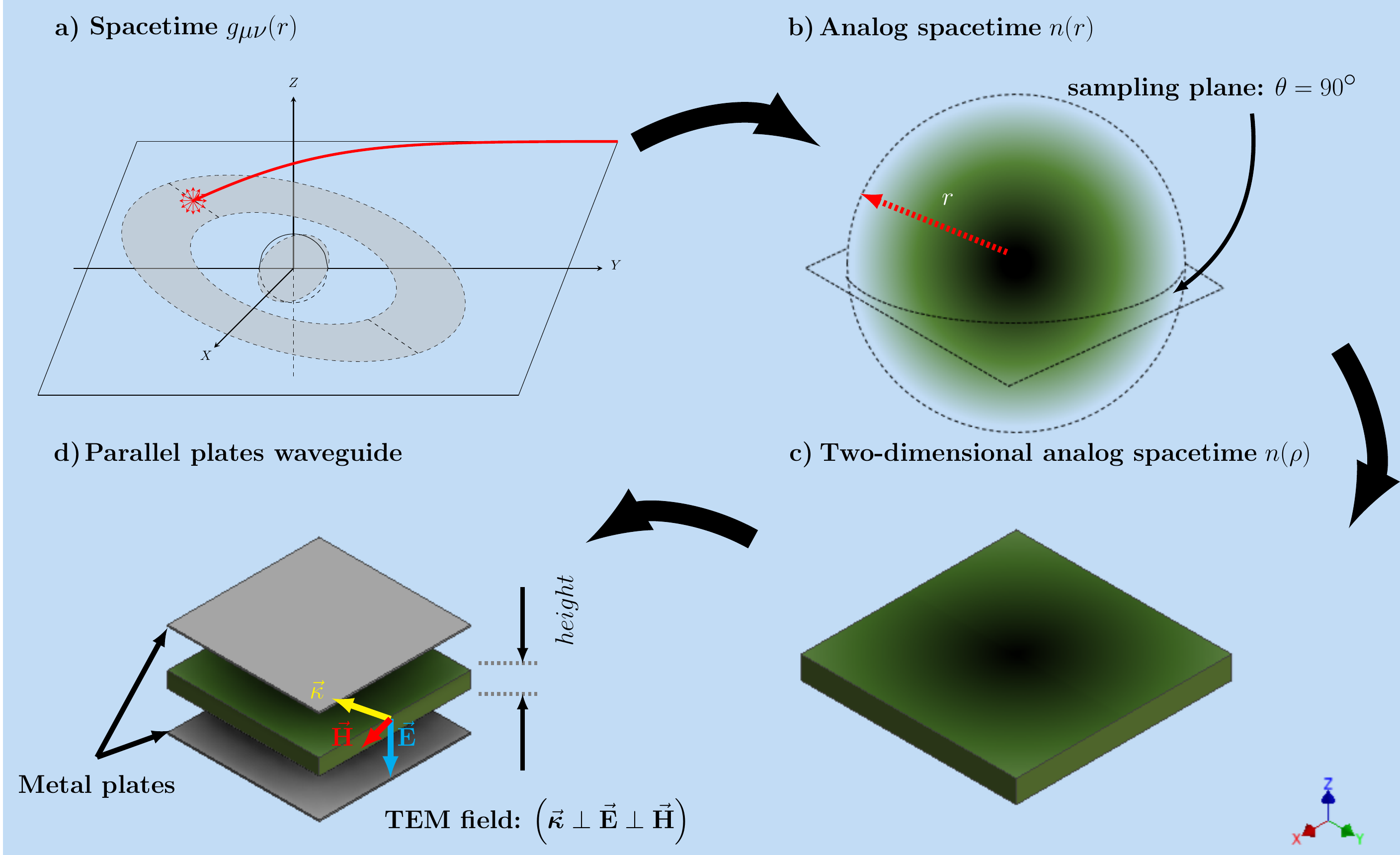}
\caption{Parallel-plate waveguide analog EM model scheme. (a) Static and spherically symmetric space-time, described by the metric $g_{\mu \nu}(r)$. (b) Analog space-time defined by the refractive index distribution $n(r)$. (c) Analog space-time of reduced dimensionality characterised by the refractive index distribution $n(\rho)$. (d) Parallel-plate waveguide analog EM model. In panel (d), the electrodynamics can be described through ray-tracing (zero wavelength approximation) in the equatorial plane, as the wave vector $\vec{\kappa}$ of the propagating TEM mode under consideration, i.e., ($\vec{{E}} = {E}_{\rm z_1}\boldsymbol{\hat z}_1$ and $\vec{{H}} = {H}_{\rho_1}\boldsymbol{\hat \rho_1}+{H}_{\varphi_1} \boldsymbol{\hat \varphi_1}$), is always parallel to the conducting plates. We checked that the ray dynamics of setups (b) in the equatorial plane and (d) are equivalent. }
\label{fig:figura_fundamentals_00}
\end{figure*}

\begin{figure*}
\centering
\includegraphics[width=18 cm]{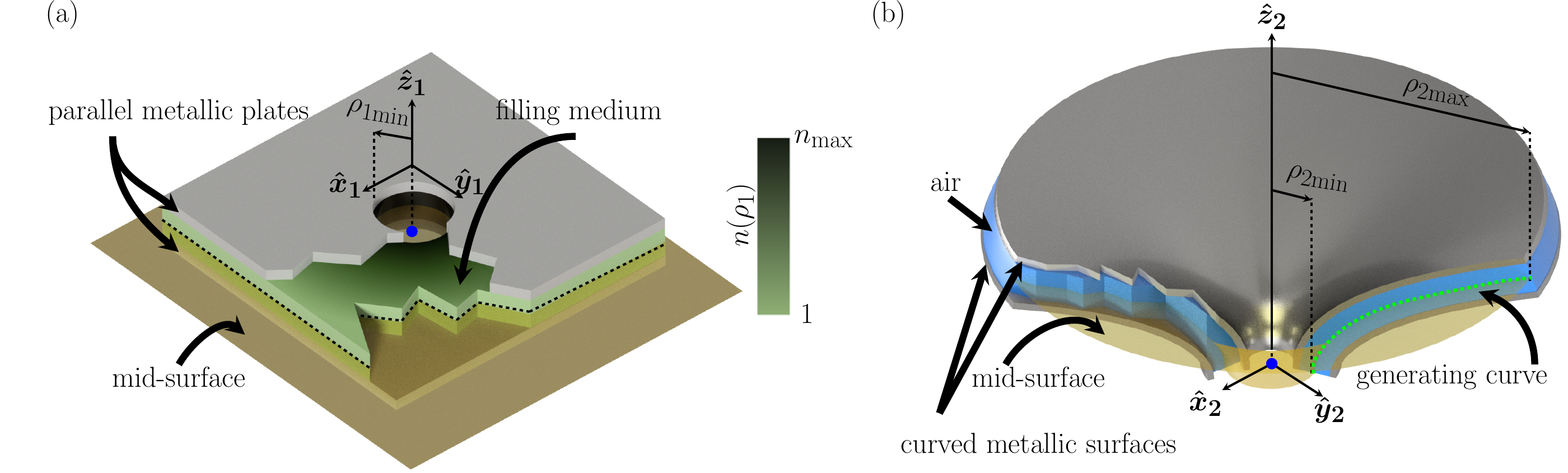}
\caption{Analog EM models of static and spherically symmetric gravitational fields. \emph{Panel (a)} shows a parallel-plate waveguide endowed with a refractive index distribution $n(\rho_1)$ with $\rho_1^2 = x_1^2+y_1^2$. \emph{Panel (b)} displays a geodesic waveguide in the interval $\rho_{2 \rm min} \leq \rho_2\leq \rho_{2 \rm max} $ with $\rho_2^2 = x_2^2+y_2^2$, where $\rho_{2 \rm min}$ is the minimum distance (with the same meaning of $\rho_{1 \rm min}$, i.e., analog event horizon or WH mouth) and $\rho_{2 \rm max}$ is the maximum radial extent of the EM device, although it theoretically extends to infinity.}
\label{fig:figura_fundamentals_2}
\end{figure*}

In the last section, the analog EM model of a static and spherically symmetric gravitational field (see Fig.~\ref{fig:figura_fundamentals_00}(a)) was represented in the form of a radially-varying isotropic refractive index distribution (see Fig.~\ref{fig:figura_fundamentals_00}(b)). Although the fabrication of a medium with such a spatial dependence added to analysing the EM wave propagation in it can be challenging, we resort to a version thereof of reduced dimensionality, which can be seen as a 2+1 analog model of gravity. In other words, the refractive index is made invariant with respect to the $z$ coordinate (see Fig.~\ref{fig:figura_fundamentals_00}(c)), thereby making it suitable to be described in cylindrical coordinates $(t_1,\rho_1,\varphi_1,z_1)$ rather than in spherical ones, namely

\begin{equation}
n = n  \left( \rho_1 \right). \label{eq:refractive_index_general_2}
\end{equation}
The Cartesian coordinates $(x_1,y_1,z_1)$ are related to the cylindrical ones, as $(t_1,x_1,y_1,z_1) = (t_1,\rho_1 \cos({\varphi}_1), \rho_1 \sin({\varphi}_1),z_1)$. We have three reasons for our choice. Firstly, in problems with cylindrical symmetry TEM field solutions of the plane wave type, e.g., modes with non-zero field components (${E}_{\rm z_1},{H}_{\rho_1},{H}_{\varphi_1}$) and (${H}_{\rm z_1},{E}_{\rho_1},{E}_{\varphi_1}$), can be contemplated thereby easing their analysis. Secondly, although the analog EM model of gravity with cylindrical symmetry \eqref{eq:refractive_index_general_2}  seems to be not electrodynamically equivalent to the refractive medium with spherical symmetry \eqref{eq:refractive_index_general}, we will show later that the paths of photons on the equatorial plane of both cases converge to the same result. Therefore, it is expected that collimated beams propagate within the waveguide following geodesics while mimicking, through the EM properties of the filling medium, other gravitational phenomena to some extent. Thirdly, a simple laboratory device can be attained by using a pair of equally-spaced metallic plates parallel to the equatorial plane, i.e., a parallel-plate waveguide, see Fig.~\ref{fig:figura_fundamentals_00}(d) and the sketch (a) in Fig.~\eqref{fig:figura_fundamentals_2}. Furthermore, in this case only the mode (${E}_{\rm z_1},{H}_{\rho_1},{H}_{\varphi_1}$) can propagate as its boundary conditions are satisfied, whereas those of the mode (${H}_{\rm z_1},{E}_{\rho_1},{E}_{\varphi_1}$) are not. 

For the moment, we assume that the flat waveguide has an infinite extent, and as it stems from the analog EM model of gravity shown in Eq.~ \eqref{eq:refractive_index_general}, the radial distance $\rho_{1\rm min}$ coincides with the size of either the analog event horizon or the analog WH mouth. In the same way, other limits of interest, such as the radius of the analog photon sphere, will be found also in the flat waveguide. However, an appropriate name for it should be analog photon circle. Furthermore, only the region $\rho_1 \geq \rho_{1\rm min}$ is represented in Fig.~\eqref{fig:figura_fundamentals_2}(a) because the EM properties of the analog EM models of gravity \eqref{eq:refraction_index_schwarzschild} and \eqref{eq:ellisRefractionIndex} comprise an exterior solution, being unspecified for $\rho_1 \leq \rho_{1\rm min}$, see Sec.~\ref{sec:grav_field}. The height of the flat waveguide is not indicated in Fig.~\ref{fig:figura_fundamentals_2}(a), since as it will be detailed later, no full-wave simulations are performed by us with this waveguide, but only the computation of ray trajectories in its mid-surface. This is necessary in the next section as a formal starting point for the issues to be addressed through the geodesic lens. Finally, we introduce the concept of \emph{mid-surface},  defined as the imaginary surface parallel to both conducting surfaces located in the middle between them.

\section{The geodesic waveguide}\label{sec:geodesic_lens_approach}
The purpose of this section is to introduce the physical principle and design guidelines of a geodesic lens, as well as the derivation of the differential equation for ray trajectories on its mid-surface. Hereafter, the geodesic lens is referred to as a geodesic waveguide. In this way, in Sec. \ref{sec:character_GL}, the features of the geodesic waveguide are described, then in Sec. \ref{sec:tailoring}, we explain how to design its mid-surface by tailoring it on the refractive index distribution within the flat waveguide. After that, the ray equation on the mid-surface is derived in Sec. \ref{sec:geodesic_equation_geodesic_waveguide}. Finally, Sec. \ref{sec:gaussian_beam} describes the one-dimensional Gaussian beam.

\subsection{Characteristics of the geodesic waveguide}
\label{sec:character_GL}
A geodesic waveguide is constructed using two parallel metallic and curved surfaces with axial symmetry and air as the filling medium, see Fig.~\ref{fig:figura_fundamentals_2}(b). We employ Cartesian coordinates $(x_2,y_2,z_2)$ to describe its geometry. In this context, ``parallel'' means that for any given point on the mid-surface, the separation between the conducting surfaces as measured along the mid-surface's normal direction is constant to $h$, while the mid-surface intersects this direction in the middle. In addition, due to the axial symmetry of the waveguide, it is convenient to employ cylindrical coordinates $(\rho_2,\varphi_2,z_2)$ to describe the mid-surface. Furthermore, the latter can be obtained by rotating a two-dimensional curve of equation $z_2=f(\rho_2)$ around the $z_2$-axis, known as the \emph{generating curve of the mid-surface}, see the dashed green profile in Fig.~\ref{fig:figura_fundamentals_2}(b) for an example.
 
Moreover, we assume that only TEM fields propagate inside the geodesic waveguide. Therefore, the minimum curvature radius $R_{\rm cmin}$ of the mid-surface must satisfy the condition $R_{\rm cmin} \gg \lambda$, so that it can be assumed to be locally flat for TEM fields, where $\lambda$ is the operating wavelength of the EM field propagating inside the waveguide.
 
\subsection{Tailoring the geodesic lens geometry}
\label{sec:tailoring}
Let us start by addressing the propagation of TEM fields in the geodesic waveguide shown in Fig.~\ref{fig:figura_fundamentals_2}(b). Our TEM-field assumption entails that the electric field is everywhere inside the waveguide locally perpendicular to the metallic surfaces, whereas the magnetic field is parallel to them. Therefore, the EM energy propagates following paths contained on the mid-surface in accordance with \emph{Fermat's principle} \cite{kunz1954propagation}. 

A coordinate system is defined on the mid-surface to model electrodynamics inside the geodesic waveguide. Let there be a local orthogonal and curvilinear coordinate system on the mid-surface, which allows us to indicate each point on it. The coordinate lines thereof are formed by intersecting the mid-surface with the coordinate surfaces of the cylindrical coordinate system. So, by intersecting the mid-surface with a plane forming an angle $\varphi_2$ with respect to the plane $\mathbf{\hat{x}}_2-\mathbf{\hat{z}}_2$, the ensuing coordinate line is given by the generating curve. The position of a point along this coordinate line is indicated by its length $S_2 = S_2(\rho_2,z_2)$ with respect to some reference and, in addition,  $\dd S_2^2 = \dd \rho_2^2+z_2^2$. On the other hand, the other type of coordinate line is obtained by intersecting the mid-surface with cylinders of radius $\rho_2$ together with planes of height $z_2$, thereby giving rise to circles of radius $\rho_2$ and height $z_2$. The position along this coordinate line is given by the angle $\varphi_2$.

So, the propagation of TEM fields can be modeled by the infinitesimal electrical length element they experience on the mid-surface \cite{kunz1954propagation}. Since the geodesic waveguide is filled with air, it coincides with the infinitesimal length element of the mid-surface
\begin{equation}
\dd L_2^2 =   \dd S_2^2 +\rho_2^2 \dd \varphi_2^2,\label{eq:interval_flat_waveguide_2}
\end{equation}

For the moment, let us consider the above infinitesimal length element only from a geometrical point of view. In the theory of differential geometry of curves and surfaces \cite{eisenhart1909treatise}, there exists an important type of spatial coordinate transformation known as \emph{ conformal coordinate transformations}, which states that \emph{if the points in two two-dimensional surfaces, namely $\Omega _1$ and $\Omega _2$, are related through a conformal transformation, the infinitesimal length elements of the surfaces, i.e., $\dd L _1^2$ and $\dd L _2^2$, will be proportional to each other via a conformal factor $K$, thus}

\begin{equation}
\dd L_1^2 = K^2 \dd L_2^2.
\end{equation}

Another geometrical interpretation for the above equation is that a local curvilinear and orthogonal coordinate system can be defined on the resulting conformal surface $\Omega _1$ in the same way as for the surface $\Omega _2$, while the angles between the ensuing coordinate lines will not change. The function of the conformal parameter is thus scaling displacements on the conformal surface $\Omega_1$ whenever a distance is measured between two points along a geodesic such that, if the same procedure is performed between equivalent points on $\Omega_2$, the same distance is obtained. Therefore, from the EM point of view, the conformal factor performs the function of a refractive index distribution. Consider the line element for TEM fields on the mid-surface of the flat waveguide; see Fig.~\ref{fig:figura_fundamentals_2}(a). We have

\begin{equation}
\dd L_1^2 = n^2(\rho_1) \left[ \dd S_1^2 +\rho_1^2 \dd \varphi_1^2\right].\label{eq:interval_flat_waveguide}
\end{equation}

We have resorted to a local coordinate system $(S_1,\varphi_1)$ on the mid-surface, $ S_1 = \rho _1$, and $\varphi_1$ corresponds to the azimuthal angle. Considering the concepts outlined above, the infinitesimal electrical length interval can be viewed as the line element of a conformal surface to that of the geodesic lens, where the refractive index distribution $n(\rho)$ is the conformal factor. The problem of the geodesic waveguide thus reduces to finding a coordinate transformation for which the effective metric tensors of the infinitesimal electrical length elements, i.e., $g_{1ij}$ (see Eq.~\eqref{eq:interval_flat_waveguide} for the flat waveguide) and $g_{2ij}$ (see Eq.~\eqref{eq:interval_flat_waveguide_2} for the geodesic waveguide),  are related via the following transformation for preserving the electrical length elements

\begin{equation}
g_{1\rm ij} = g_{2ij}\frac{\partial x_2^i}{\partial x_1^i}\frac{\partial x_2^j}{\partial x_1^j },\label{eq:tensor_equation_flat_geo}
 \end{equation}
where $x^{\rm i}_{\rm k}$ are the curvilinear and orthogonal coordinates either on the mid-surface of the flat ($k = 1$) or the geodesic waveguide ($k = 2$). Therefore, the conditions for Eq.~\eqref{eq:tensor_equation_flat_geo} to be fulfilled yield the following differential equations and algebraic relationships
\begin{subequations}
\begin{align}
n(\rho_1)  &=  \dfrac{\dd S_2}{\dd S_1},\label{eq:z_relatinoship_A}\\
n(\rho_1) \rho _1 &=  \rho _2,\label{eq:z_relatinoship_B}\\
\varphi_1 &=  \varphi _2.\label{eq:z_relatinoship_C}
\end{align}
\end{subequations}
As $\dd S_2 ^2 = \dd \rho_2^2+\dd z_2^2$ and $\dd S_1 = \dd \rho_1$, Eq.~\eqref{eq:z_relatinoship_A} assumes the following form
\begin{equation}
n^2(\rho_1) \dd \rho_1^2 = \dd \rho_2^2+\dd z_2^2.\label{eq:z_relatinoship_BB}
\end{equation}
Substituting $\dd \rho_1/\dd \rho_2$ from Eq.~\eqref{eq:z_relatinoship_B}, the differential equation for the generating curve of the mid-surface is
\begin{equation}
\left( \frac{\dd z_2}{\dd \rho_2}\right)^2 = \frac{n^2(\rho_1)}{[n(\rho_1)+n'(\rho_1)\rho_1]^2} -1,\label{eq:z_relatinoship_CC}
\end{equation}
where $n'(\rho_1) = \dd n(\rho_1)/\dd \rho_1$. However, it is convenient to parametrize Eq.~\eqref{eq:z_relatinoship_CC} as a function of $\rho_1$, since the right-hand side is a function thereof. To this end, we use Eq.~\eqref{eq:z_relatinoship_B} to obtain $\dd \rho_2/\dd \rho_1$, and then
\begin{align}
\left( \frac{\dd z_2}{\dd \rho_1}\right)^2 &= -\rho_1n'(\rho_1)\Biggr{[}  2n(\rho_1)+n'(\rho_1)\rho_1\Biggr{]}.\label{eq:z_relatinoship_D}
\end{align}
We note that the right-hand side of Eq. \eqref{eq:z_relatinoship_D} must be positive. Therefore, the space-time geometry encoded into the refractive index distribution in the flat waveguide might not have a representation as a geodesic lens to its full extent. Then, the flat waveguide can be represented as a geodesic waveguide in the domain, where
\begin{equation}
\frac{n'(\rho_1) \rho _1}{2n(\rho_1)}\geq -1. \label{eq:z_relatinoship_E}
\end{equation}

\subsection{Ray trajectories at the geodesic waveguide's mid-surface}
\label{sec:geodesic_equation_geodesic_waveguide}
Using Lagrangian optics \cite{lakshminarayanan2002optical,gomez2016three}, we obtain the differential equation governing the ray trajectories at the mid-surface of the geodesic waveguide. Apart from that, we also demonstrate that such geodesics are equivalent to those given in the flat waveguide framework and, simultaneously, to the exact solutions of GR.

The Lagrangian on the mid-surface of the geodesic waveguide is derived from Eq. \eqref{eq:interval_flat_waveguide_2} as
\begin{equation}
\mathcal{L} =\sqrt{g_{2ij}\frac{\dd x_2^i}{\dd S_2}\frac{\dd x_2^j}{\dd S_2}}= \sqrt{1+\rho_2 ^2 \varphi_{2S_2}^2},
\end{equation}
where $\varphi_{S_2} = \dd \varphi _2/\dd S_2$. As the Lagrangian does not depend on $\varphi_2$, then there exists the conserved angular momentum $L_{\varphi_2}=\partial \mathcal{L}/\partial \varphi_{2S_2}$ \cite{hamill2014student,lakshminarayanan2002optical,gomez2016three}, which leads to the ray equation on the geodesic waveguide's mid-surface:
\begin{equation} \label{eq:diff_eq}
\left(\frac{\dd S_2}{\dd \varphi_2}\right)^2 = \rho_2^4 \left(\frac{1}{L_{\varphi_2}^2}-\frac{1}{\rho_2^2} \right).
\end{equation}
Note that the above equation is parametrized in terms of the variable $\varphi_2$ for the sake of simplicity. By using the same procedure on the flat waveguide, we obtain the ray equation on its mid-surface
\begin{equation}
 \left(\frac{\dd \rho_1}{\dd \varphi_1}\right)^2 = n^2(\rho_1)\rho_1^4 \left[\frac{1}{L_{\varphi_1}^2}-\frac{1}{n^2(\rho_1)\rho_1^2} \right].\label{eq:ray_equation}
\end{equation}

It can be easily demonstrated that both Eqs.~\eqref{eq:diff_eq} and ~\eqref{eq:ray_equation} are equivalent by substituting $\dd S_2^2 = \dd \rho_2^2+ \dd z_2^2$, and $L_{\varphi_2} = L_{\varphi_1}$  into Eq.~\eqref{eq:diff_eq} and using the transformation relationships \eqref{eq:z_relatinoship_A}, \eqref{eq:z_relatinoship_B}, and \eqref{eq:z_relatinoship_C}. Moreover, Eq.~\eqref{eq:ray_equation} corresponds also to the null-geodesic equation in the original gravitational framework described by Eq.~\eqref{eq:isotropic_metric_tensor}, as it can be easily verified via the Lagrangian approach \cite{falcon2022propagation}.
\subsection{The one-dimensional Gaussian beam}\label{sec:gaussian_beam}
In this section, we introduce the definition and some relevant properties of a one-dimensional Gaussian beam, from now on just called Gaussian beam for shortness.

A Gaussian beam is a flux of radiation, which is very similar to an EM plane wave due to the following two main aspects: (1) it has a clearly defined propagation direction; (2) both electric and magnetic field components are mutually perpendicular between them, and in turn, also to the propagation direction \cite{goldsmith1998quasioptical}. Instead, the fundamental difference relies on the Gaussian beam amplitude, which varies and asymptotically decreases along directions perpendicular to the propagation axis. 

Let us consider an air-filled parallel-plate waveguide similar to that depicted in Fig.~\ref{fig:figura_fundamentals_2}(a), where a Gaussian beam propagates along the $x_1$-axis, and its amplitude only varies along the $y_1$-axis. The electric field inside the waveguide satisfies the paraxial wave equation \cite{goldsmith1998quasioptical} (we set $(x_1,y_1,z_1)=(x,y,z)$ to lighten the notation):
\begin{equation}
\frac{\partial^2 E_z(x,y)}{\partial  y^2}-2j\kappa \frac{\partial E_z(x,y)}{\partial  x} = 0,\label{eq:helmholtz_equation}
\end{equation}
where $\kappa$ is the wave number and $E_z$ is the $z$ component of the electric field. The solution of Eq. \eqref{eq:helmholtz_equation} is given by the well-known formula (see pp. 16 in Ref. \cite{goldsmith1998quasioptical}) 
\begin{align}
E_z(x,y) &= \frac{E_0}{\sqrt{w(x)}}\exp\left[-\left( \frac{y}{w(x)}\right)^2-j \kappa x\right.\notag\\
&\left.-j\frac{\kappa y^2}{2 R(x)}+\Phi_0(x)\right],\label{eq:gaussian_beam}
\end{align}
with $E_{0}$ being the field amplitude, $w(x)$ the beam radius, $R(x)$ the radius of curvature of the wavefront, and $\Phi_0(x)$ the Gaussian phase shift \cite{goldsmith1998quasioptical}, whose expressions are 
\begin{subequations}
\begin{align}
R(x) &= x+\frac{1}{x}\left ( \frac{\pi w_0^2}{\lambda}\right )^2,\label{eq:beam_radius_N}\\
w(x) &= w_0\sqrt{1+\left ( \frac{\lambda x}{\pi w_0^2}\right )^2},\label{eq:beam_waist_N}\\
\Phi_0(x) &= \tan^{-1} \left( \frac{\lambda x}{\pi w_0^2} \right),
\end{align}
\end{subequations}
where $w_0$ is the waist (minimum beam radius). The point where the waist \eqref{eq:beam_waist_N} of the Gaussian beam is minimum is called beam waist position, i.e., $\vec{r}=(0,0)$ according to Eq.~\eqref{eq:beam_waist_N}. In this case, Eq.~\eqref{eq:beam_waist_N} models how the beamwidth varies along the propagation axis. Furthermore, the beamwidth increases by a factor of $\sqrt{2}$ relative to its minimum value $w_0$ at the confocal distance given by $x_{\rm con} = w_0^2 \pi /\lambda$.

\section{Results}
\label{sec:results}
In this section, the results of the geodesic waveguide mimicking the effects exerted on EM waves by both a Schwarzschild BH (see Sec. \ref{sec:results_shcwarzschild_BH}) and a Morris-Thorne WH (see Sec. \ref{sec:results_Morris_Thorne_WH}) are presented and then discussed. Eventually, a comparison of our results with the state-of-the-art literature is addressed in Sec.~\ref{sec:SoA}. 

\begin{figure}[H]
\begin{center}
\includegraphics[width = 8cm]{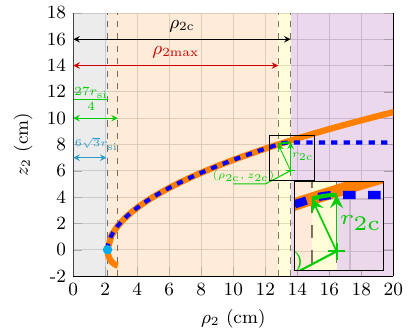}
\end{center}
\vspace{-0.8cm}
\caption{Generating curve of the geodesic waveguide for a Schwarzschild BH ($r_{\rm si1} =2.05~\rm mm$) represented by a solid orange line, while the dashed blue line is its truncated version. The inset plots show the smooth transition region, which is modeled by a circular arc.}\label{fig:z_2_vs_rho_2_BH}
\end{figure}
\subsection{The analog geodesic waveguide of the Schwarzschild BH}\label{sec:results_shcwarzschild_BH}
We first describe the technical specifics to build up the geodesic waveguide (see Sec. \ref{sec:specifics}) and then provide the numerical simulations (see Sec. \ref{sec:NS}).

\subsubsection{Technical specifics to build up the geodesic lens }
\label{sec:specifics}
First, it is important to recall that, as explained in Sec.~\ref{sec:geodesic_lens_approach}.B, the shape of each geodesic waveguide is obtained by applying a conformal coordinate transformation to the points on the mid-surface of a flat waveguide outlined in Sec.~\ref{sec_the_analog_em_model_of_gravity}.C. Thus, it will be necessary to handle variables in both domains in order to attain the geodesic waveguide. Coordinates and variables with a subscript $\rm i = 1$ are referred to the flat waveguide's domain, whereas those with subscript $\rm i = 2$ are referred to the geodesic waveguide's domain. Furthermore, this distinction is also important because the effective trajectories followed by the Gaussian beams in the geodesic waveguide will be compared with the trajectories obtained from the ray equation \eqref{eq:ray_equation} framed on the mid-surface of the flat waveguide.

The carrier frequency of the Gaussian beam exploited in our tests is set to $f_{\rm c} = 140~{\rm GHz}$ (or wavelength $\lambda = 2.1~\rm mm$). We use as analog Schwarzschild radius $r_{\rm si1} = 2.05~\rm mm$. Thus, as will be shown later, the radius of the analog photon circle satisfies the condition $\rho_{2}\geq 10 \lambda$, see Sec.~\ref{sec:character_GL}. Then, considering the refractive index distribution shown in Eq.~\eqref{eq:refraction_index_schwarzschild}, we use \texttt{MATLAB R2022b} \cite{mathworks2023matlab} to solve Eqs.~\eqref{eq:z_relatinoship_B} and \eqref{eq:z_relatinoship_D} for determining the generating curve of the mid-surface, see orange line in Fig.~\ref{fig:z_2_vs_rho_2_BH}. In particular, the inequality \eqref{eq:z_relatinoship_E} is not satisfied in the interval $r_{\rm si1} \leq \rho_1 \leq  2r_{\rm si1} $. Therefore, the orange line of Fig.~\ref{fig:z_2_vs_rho_2_BH} encodes only the effective metric induced by the refractive medium \eqref{eq:refraction_index_schwarzschild} within the range $\rho_1 \geq  2r_{\rm si} $ (excluding the analog event horizon).

The generating curve can be divided into two branches ($z_2 <0$ and $z_2 >0$), both joining at the cyan point $(\rho_2,z_2) = (6 \sqrt{3}r_{\rm si1},0)$. Here $\rho_{2} = 6 \sqrt{3} r_{\rm si1}$ corresponds to the photon circle radius \cite{falcon2022propagation} as projected on the mid-surface of the geodesic waveguide via Eq.~\eqref{eq:z_relatinoship_B}, see Appendix \ref{sec:photon_sphere} for details. We choose the positive branch to construct the geodesic waveguide because its inner limit corresponds to the analog photon circle. The latter is a suitable location to place an absorbing boundary condition to emulate the absorption effects of an analog BH (once the waveguide is implemented as a three-dimensional structure). Likewise, the generating curve may be implemented to its whole extent. However, the absorbing boundary condition has to be placed at the open end of the waveguide corresponding to the bottom loose end of the generating curve shown in Fig.~\ref{fig:z_2_vs_rho_2_BH}. This is commented on in further detail later. The generating curve is truncated at $\rho_2 = \rho_{2 \rm max}$ (see the segment of the dashed blue line within the orange region in Fig.~\ref{fig:z_2_vs_rho_2_BH}). We use $\rho_{2 \rm max}\leq x_{\rm con}$ {(see the confocal distance $x_{\rm con}$ in Sec.~\ref{sec:gaussian_beam})} to ensure that the Gaussian beam remains collimated in the region of interest inside the geodesic waveguide, that is $\rho_2 \leq \rho_{2 \rm max}$, where  $\rho_{2 \rm max} = 12.80~\rm cm$ and $x_{\rm con} = 14.66~\rm cm$.

Next, the curved blue line depicted in Fig.~\ref{fig:z_2_vs_rho_2_BH} transitions smoothly into a horizontal line within the violet-shaded region. The transition curve, illustrated by the green segment in the inset in Fig.~\ref{fig:z_2_vs_rho_2_BH}, is defined as a segment of a circle with a radius $r_{2\rm c} = 10\lambda$ and a center at $(\rho_{2\rm c},z_{2 \rm c})=(13.58~\rm cm,6.01~\rm cm)$. The coordinates $\rho_{2\rm c}$ and $z_{2 \rm c}$ are determined based on the criterion of ensuring continuity of derivatives along the profile of the generating curve.
\begin{figure}
\centering
\includegraphics[width=8 cm]{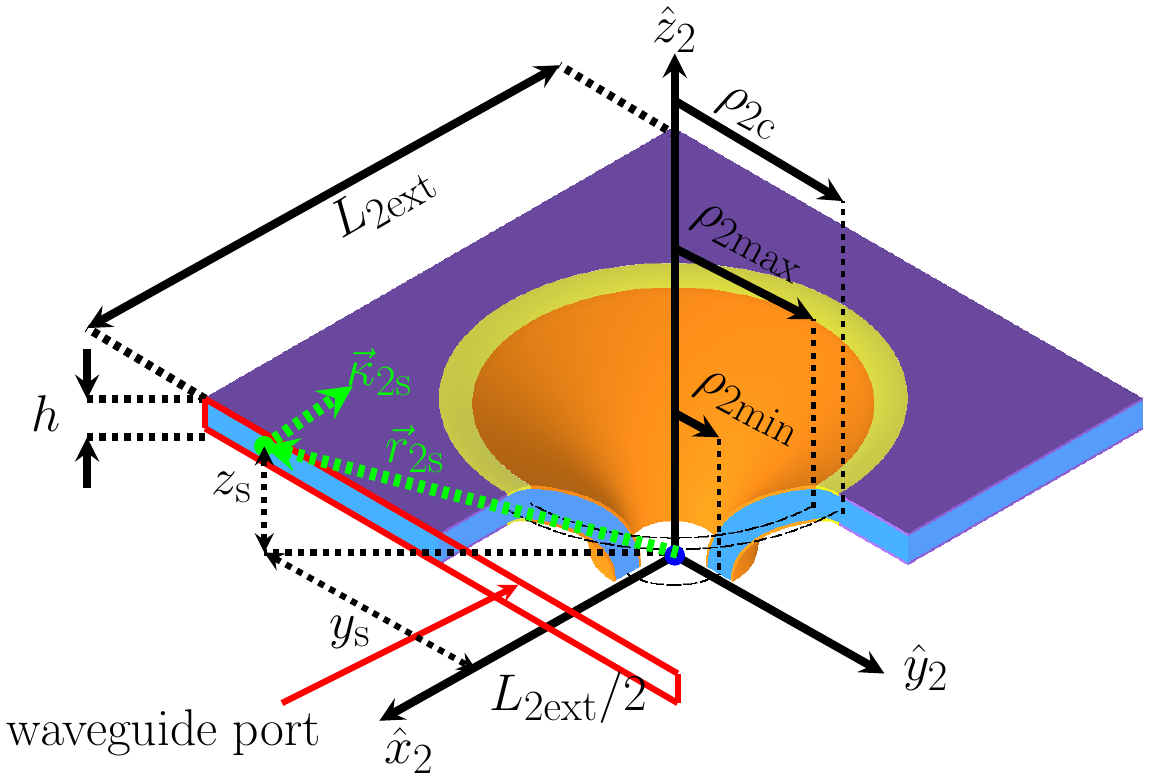}
\caption{A sectional view of the air-filled geodesic waveguide. The conducting surfaces are divided into three regions shaded in accordance with the color palette used to identify each zone of the generating curve of the mid-surface shown in Fig.~\ref{fig:z_2_vs_rho_2_BH}. The height $h$ is exaggerated for the sake of visualization. This figure is realized with the program \texttt{Inventor}\textregistered\     \cite{autodesk2023inventor}.}
\label{fig:figura_fundamentals_3}
\end{figure}
A three-dimensional model of the geodesic waveguide is presented in Fig.~\ref{fig:figura_fundamentals_3}. The separation between the conducting surfaces is $h = 1~\rm mm$ ($h\leq \lambda/2$ for a TEM parallel-plate waveguide, see pp. 102 in Ref. \cite{pozar2011microwave}). In addition, the extent of the waveguide is limited within a square of side $ L_{2 \rm ext} = 2 (\rho_{2\rm c}+d)$. We have considered a distance of $d = 15 \lambda$ to ensure a sufficiently large area for the establishment of a TEM mode.

Subsequently, the model is imported into the finite element method-based solver of High-Frequency Structure Simulator (HFSS) 2021a\textregistered\  \cite{ansys2023hfss}. We define the top and bottom surfaces of the geodesic waveguide (orange, yellow, and violet-colored faces) as perfect electric conductors. Meanwhile, the remaining faces, including that in contact with the $x_2-y_2-$plane, are configured as absorbing boundary conditions. It is worth mentioning that the absorbing boundary condition on the $x_2-y_2-$plane accounts for the effect of EM radiation propagating towards the analog event horizon, which, as mentioned earlier, cannot be modeled by the geodesic waveguide. The filling medium is assumed to be air.
  As for the excitation of the geodesic waveguide, this is done through the open face enclosed by the red rectangle in Fig.~\ref{fig:figura_fundamentals_3}, which is designated as the waveguide port. 
 The excitation consists of an incident wave condition set as a Gaussian beam at the waveguide port with polarization along the $z_2$-axis, carrier frequency $f_{\rm c}$, waist $w_0 =10~\rm mm$, wave vector $\vec{\kappa}_{2 \rm s} = -\boldsymbol{\hat{x}_2}$, and beam emission position (where the beam radius is minimum and beam amplitude maximum) $ \vec{r}_{2 \rm s} = (L_{2 \rm ext}/2,y_{\rm s}, z_{\rm s})$, see $\vec{r}_{2 \rm s}$ in Fig.~\ref{fig:figura_fundamentals_3}. In our numerical experiments, the coordinate $y_{\rm s}$ is varied in the interval $-L_{2 \rm ext}/2 \leq y_s \leq L_{2 \rm ext}/2$ while $z_{\rm s}$  is set to the maximum height of the mid-surface. Moreover, a Gaussian beam, as launched by HFSS, exhibits a variation in its amplitude along both $y_2$ and $z_2$ directions. However, upon propagating inside the violet region in Fig.~\ref{fig:figura_fundamentals_3}, it is expected to excite a Gaussian beam with the characteristics discussed in Sec.~\ref{sec:gaussian_beam}. Besides this, for implementation purposes, a Gaussian beam launcher \cite{tobia2018design} can be designed to be used together with other standard feeding interfaces (e.g., a rectangular waveguide \cite{pozar2011microwave}).
 \begin{figure*}[hbtp!]
\centering
\includegraphics[width=16 cm]{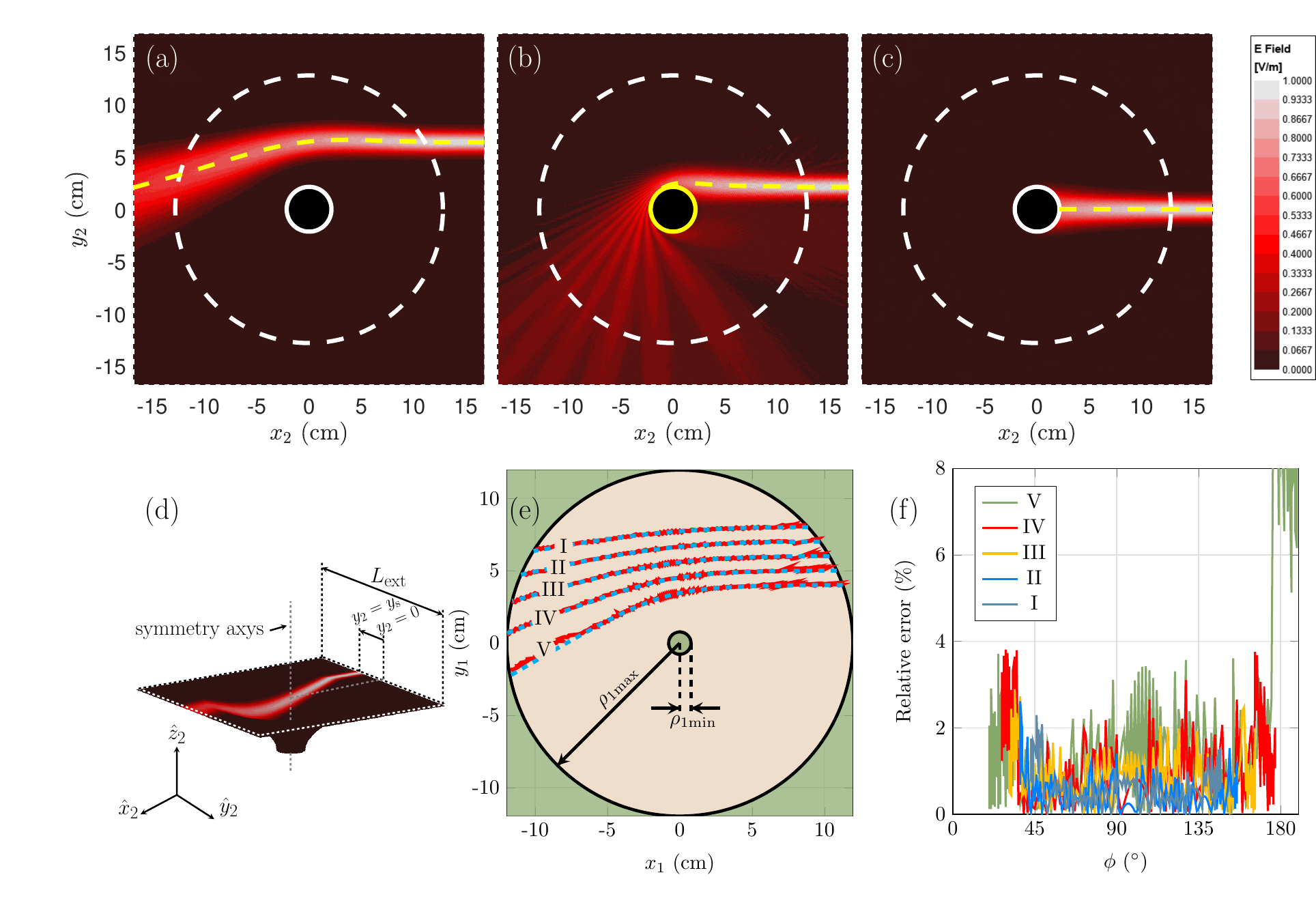}
\vspace{-0.5cm}
\caption{Results for the propagation of a Gaussian beam on the mid-surface of the geodesic waveguide for a Schwarzschild BH. \emph{Panels (a), (b), and (c).} Normalized distribution of the complex amplitude of the electric field (color map) for  $y_{\rm s} = 30,\ 10,\ 0\ \lambda$, respectively. Dashed yellow lines are the ray trajectories (corresponding to those of GR). The solid and dashed white circles delimit the photon circle and the inner region border, respectively. \emph{Panel (d).} Perspective view of the electric field distribution on the mean surface for $y_{\rm s} = 30 \lambda$. \emph{Panel (e).} Comparative image (in isotropic coordinates) between the exact solutions from GR (cyan lines) and the approximated trajectories of the Gaussian beam (solid red lines) for $y_{\rm s} = 40\lambda\ (\rm{I}),\ 35\lambda\ (\rm{II}),\ 30\lambda\ (\rm{III}),\ 25\lambda\ (\rm{IV}),\ 20 \lambda\ (\rm{V})$. \emph{Panel (f).} Relative error along the null geodesic trajectories of panel (e).}
\label{fig:result_BH_A}
\end{figure*}

As we will discuss later, one of our numerical experiments to assess the performance of the geodesic waveguide as an analog EM model of gravity involves two procedures: Firstly, computing the effective path that a Gaussian beam follows on the mid-surface of the geodesic waveguide. Secondly, this trajectory is then projected onto the mid-surface of the flat waveguide in order to compare it with a theoretical ray trajectory, which is obtained through the ray equation under appropriate initial conditions. Therefore, modifications implemented on the generating curve in the interval $\rho_2 \geq \rho_{2 \rm max}$ entail that the refractive index distribution of the corresponding flat waveguide, see Fig.~\ref{fig:figura_fundamentals_2}(a), must be calibrated for the region $\rho_1 \geq \rho_{1 \rm max}$ in this domain, which corresponds to the range $\rho_2 \geq \rho_{2 \rm max}$. To this end, we first solve the implicit Eq.~\eqref{eq:z_relatinoship_B} to obtain $\rho_1=\rho_1(\rho_2)$. Then,  $n'$ is solved from Eq.~\eqref{eq:z_relatinoship_C} as
\begin{subequations}
\begin{align}
n'(\rho_2)  &= \left\{ \frac{n(\rho_1)}{\rho_1} \left[ \frac{1-K(\rho_2)}{K(\rho_2)}\right]\right\}_{\rho_1 =\rho_1(\rho_2)},\label{eq:z_relatinoship_DC}\\
K(\rho_2) &= \sqrt{1+\left( \frac{\dd z_2}{\dd \rho_2} \right)_{\rho_1 =\rho_1(\rho_2)}^2}\ .
\end{align}    
\end{subequations}
After some calculations, according to the geometry of the generating curve the factor $K$ in the yellow and violet regions in Fig.~\ref{fig:figura_fundamentals_3} turns out to be
\begin{equation}
  K(\rho_2)=\begin{cases}
    \dfrac{r_{\rm 2c}}{\sqrt{r_{\rm 2c}^2-(\rho_2-\rho_{2 \rm c})^2}}, & \text{if $\rho_{2 \rm max} \leq \rho_2 \leq \rho_{2 \rm c}$}.\\
    1, & \text{if $ \rho_2 \geq \rho_{2 \rm c}$}.
  \end{cases}\label{eq:cases_K}
\end{equation}
The relative error between the calibrated and uncalibrated refractive index distribution is below $10^{-2}$, and Eq.~\eqref{eq:z_relatinoship_B} remains still true.

\subsubsection{Numerical simulations}
\label{sec:NS}
The results shown in Figs.~\ref{fig:result_BH_A}(a), (b), and (c) display the normalized distribution of the complex amplitude of the electric field on the mid-surface for three distinct cases of the beam emission position at the waveguide port: $y_{\rm s}= 30\lambda,10\lambda,0$ (top view perspective). The region of interest comprises the region demarcated by the dashed and solid white circles, whose radii are those of the analog photon circle and $\rho_{2 \rm max}$, respectively. Figure~\ref{fig:result_BH_A}(d) shows the perspective view of the mid-surface as the color map displays the complex amplitude of the electric field shown in panel (a). To compute the exact ray trajectories for each scenario, as shown by the dashed yellow paths, we use as initial conditions the Gaussian beam emission position and its wave vector on the waveguide port, i.e., $\vec{r}_{2 \rm s}$ and $\vec{\kappa}_{2 \rm s}$ as projected on the flat waveguide's mid-surface through the coordinate transformation~\eqref{eq:z_relatinoship_B}. The angular momentum, a necessary parameter to solve the ray equation, is given by \cite{Defalco2016}
\begin{equation}
L_{\phi} = \rho_{20} \sin\alpha,
\end{equation}
where $\rho_{20} = \vec{r}_{2 \rm s} \cdot \boldsymbol{\hat{\rho}}_2$ represents the initial radius and $\alpha$ the emission angle, given by $\cos\alpha =  \vec{\kappa}_{2 \rm s} \cdot \boldsymbol{\hat{\rho}}_2/ |\vec{\kappa}_{2 \rm s}|$ (applicable only within the violet region in Fig.~\ref{fig:figura_fundamentals_3}).
We note that in panel (a), the beam spreads toward the edge of the waveguide, and its effective path agrees very well with the yellow dashed trajectory. In panel (c), the beam propagates towards the photon sphere, where the absorbing boundary condition completely absorbs it. Finally, in panel (b), the passage of the beam near the analog photon sphere, which is an unstable region for ray trajectories \cite{falcon2022propagation}, entails that part of it is lost due to the absorbing boundary condition placed at the photon circle (inner extreme of the waveguide); whereas another part is redistributed within the waveguide. Although the Gaussian beam splits into several parts, the yellow-dashed path has a good agreement with the Gaussian beam's effective trajectory before the latter reaches the photon sphere. 

We now analyze the phenomenon as observed on the mid-surface of the flat waveguide. Figure~\ref{fig:result_BH_A}(e) illustrates the trajectories followed by the Gaussian beam projected onto the mid-surface of the flat waveguide (depicted as solid red lines). Here, both $\rho_{1 \rm min}$ and $\rho_{1 \rm max}$ correspond to the photon circle size and $\rho_{2 \rm max}$, respectively. Each case is labelled by I, II, III, IV, and V, for initial conditions $y_{\rm s} = 40\lambda,35\lambda,30\lambda,25\lambda,20\lambda$, respectively. These trajectories are obtained by considering the ridge of the electric field amplitude distribution on the mid-surface of the geodesic waveguide and subsequently projecting them onto the mid-surface of the corresponding flat waveguide. The cyan dashed lines represent the solutions of the ray equation \eqref{eq:ray_equation}, computed under the appropriate initial conditions. The trajectories followed by the Gaussian beam $\rho_{\rm beam}(\phi)$ in cases I -- IV converge to those obtained using the ray equation $\rho_{\rm ray}(\phi)$, because the relative error in the radial coordinate  
\begin{equation}
\text{Error}_{\text{rel}}(\phi) = \frac{|\rho_{\rm beam}(\phi)-\rho_{\rm ray}(\phi)|}{\rho_{\rm ray}(\phi)}\cdot 100 \left[\%\right] ,
\end{equation}
remains below 4\%, see Fig.~\ref{fig:result_BH_A}(f). However, case V shows an apparent disagreement near the trajectory's left end as its relative error reaches values as large as 8\% . This might be due to the inaccuracies in the method employed for extracting the ridge from the complex field distribution on the geodesic waveguide's mid-surface.

\subsection{The analog geodesic waveguide of the Morris-Thorne WH}
\label{sec:results_Morris_Thorne_WH}
\begin{figure*}[hbtp!]
\centering
\includegraphics[width = 13.5 cm]{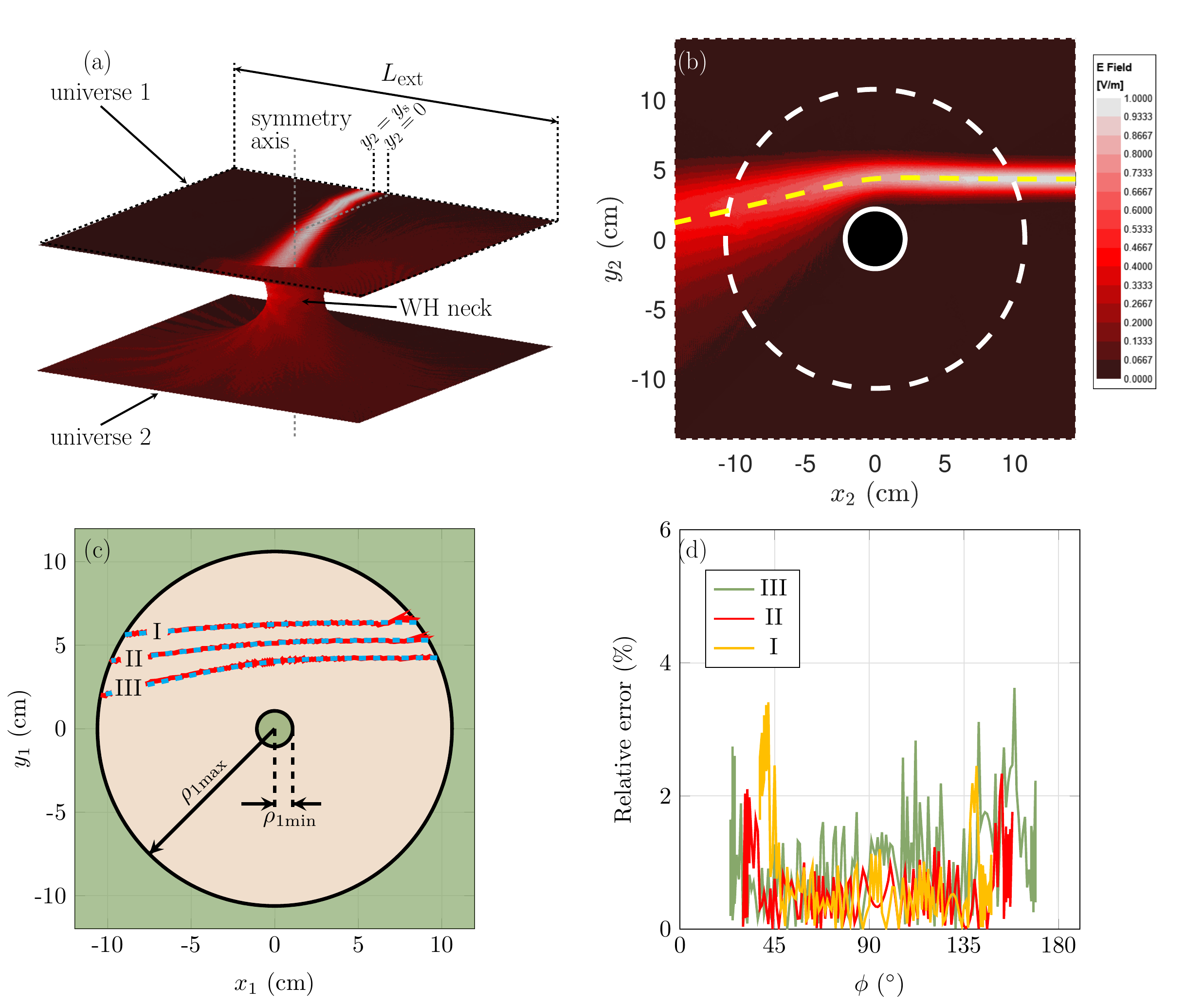}
\caption{Results for the propagation of a Gaussian beam in the EM analog model of a Morris-Thorne WH. \emph{Panel (a).} Perspective view of the distribution of the electric field's complex amplitude on the mid-surface for initial condition $y_{\rm s} = 10 \lambda$. \emph{Panel (b).} Distribution of the complex amplitude of the electric field in the universe 1 for a Gaussian beam emitted with initial condition $y_{\rm s} = 20 \lambda$. \emph{Panel (c).} Universe 1 in the flat waveguide framework. Comparison between exact solutions from GR (dashed cyan lines) and approximated trajectories followed by the Gaussian beam (solid red lines) for $y_{\rm s} = 30\lambda\ (\rm{I}),\ 25\lambda\  (\rm{II}),\ 20\lambda\ (\rm{III})$. \emph{Panel (d).} Relative error related to panel (c).}
\label{fig:result_WH_A}
\end{figure*}
We now concentrate on the Morris-Thorne WH, whose analog EM model of gravity is based on Eq.~\eqref{eq:ellisRefractionIndex}. Similarly to the analog BH case addressed in the preceding section, a flat waveguide model corresponds to the geodesic waveguide to be designed in this section. We set $b_{\rm oi1} = 1.07~\rm cm$, and the generating curve of the mid-surface is obtained from Eqs.~\eqref{eq:z_relatinoship_B} and \eqref{eq:z_relatinoship_D}, whose analytical solution is:

\begin{equation}
{z_2}({\rho}_2) = \pm 2b_{\rm oi1} \log{ \left(\frac{{\rho}_2 +\sqrt{{\rho}_2 ^2-4 b_{\rm oi1}^2}}{2b_{\rm oi1}}\right)},\ \text{for }{\rho}_2 \geq 2b_{\rm oi1}.
\label{eq:Ellis_WH_generating_curve}
\end{equation}
We note that the above solution coincides with that determined by He and collaborators \cite{he2020simulation}, who worked in the optical domain with a fully dielectric geodesic lens. 

Again, the resulting generating curve can be divided into two branches: universe 1 ($z_2 > 0$) and universe 2 ($z_2 < 0$), both spanning in the interval $\rho_2 \geq  2b_{\rm oi1}$, with $2b_{\rm oi1}$ being the size of the analog photon sphere in the geodesic waveguide frame. Similarly, as in Sec. \ref{sec:specifics}, the generating curve is likewise truncated and smoothly attached to a flat segment. In Table \ref{table:simulation_parameters}, the structural parameters of the geodesic waveguide after truncation are displayed.
\begin{table}[!h]
\caption{Parameters of the geodesic waveguide structure for mimicking the Morris-Thorne WH.}
\centering
\begin{tabular}{|c|c|}
\hline
   \hspace{0.2cm}{\bf Parameter}\hspace{0.2cm}  &\hspace{0.2cm} {\bf Value}\hspace{0.2cm} \\ \hline
   $\rho_{2\rm max}$ &  10.71 cm\\  \hline
   $\rho_{\rm  2c}$ & 11.14 cm \\  \hline
   $z_{\rm  2c}$ &2.81 cm\\  \hline
   $r_{\rm  2c}$ &2.14 cm\\  \hline
   $L_{\rm 2 ext}$ &$2(\rho_{2\rm c}+15 \lambda)$\\  \hline
\end{tabular}\label{table:simulation_parameters}
\end{table}

Likewise, the refractive index distribution of the flat waveguide modeling the Morris-Thorne WH, see Eq.~\eqref{eq:ellisRefractionIndex}, is calibrated to account for the modifications introduced on the generating curve of the geodesic waveguide; thereby obtaining an absolute difference between the uncalibrated and calibrated effective refractive index distributions remains below $10^{-2}$.

This geodesic waveguide presents a similar structure to that of the Schwarzschild BH, with the exception that it is mirrored with respect to the $x_2-y_2$-plane, see for example the mid-surface depicted in Fig.~\ref{fig:result_WH_A}(a). The color map in the figure represents the distribution of the complex amplitude of the electric field. The Gaussian beam is emitted from the universe 1 with $y_{\rm s} = 10 \lambda$. In this case, part of the beam passes through the WH mouth and goes to Universe 2, while another part remains and redistributes in Universe 1. In general, the propagation dynamics of a Gaussian beam in this geodesic waveguide is similar to that obtained in the analog EM model of the Schwarzschild BH. For example, as shown in Fig.~\ref{fig:result_WH_A}(b) for $y_{\rm s} = 20 \lambda$, the apparent trajectory followed by the Gaussian beam on the universe 1 agrees with the yellow dashed trajectory, which is obtained on the mid-surface of the flat waveguide through the ray equation \eqref{eq:ray_equation} and later projected on the mid-surface of the geodesic waveguide. 

Finally, Fig.~\ref{fig:result_WH_A}(c) displays other cases of ray trajectories in the flat waveguide framework (Universe 1), having as initial conditions $y_{\rm s} = 30 \lambda,~25 \lambda,~20 \lambda$. The flat waveguide framework, corresponding to Universe 2 is not displayed in Fig.~\ref{fig:result_WH_A}(c) because no energy passed through the WH mouth for the above-mentioned initial conditions. We see that the ray trajectories followed by the Gaussian beam on the mid-surface of the geodesic waveguide (solid red lines) agree with those obtained from the ray equation \eqref{eq:ray_equation} (cyan lines). In addition, the relative error measured in the radial coordinate for each case is below 4\% as shown in Fig.~\ref{fig:result_WH_A}(d).

\subsection{Comparison to the state-of-the-art solutions}
\label{sec:SoA}
The dynamics of Gaussian beams displayed in this article resembles those obtained by Fernandez \cite{fernandez2016anisotropic}, He et al. \cite{he2020simulation}, Sheng \cite{sheng2013trapping}, and Cheng \cite{cheng2010omnidirectional}. However, apart from Fernandez, who simulated an analog EM model of a Schwarzschild BH based on metamaterials and whose effective EM properties fulfill the space-time consistency conditions, other studies have not addressed this issue, as they rely on a TM mode propagating in a dielectric layer\cite{he2020simulation,sheng2013trapping,schultheiss2010optics}. Thus, as the geodesic waveguide proposed in this work operates with a TEM mode propagating within the air medium between the curved metallic surfaces, the space-time consistency conditions are naturally fulfilled. 

An interesting feature of our approach is that, unlike other analog EM models proposed in the microwave frequency range, there is no need for metamaterials or even a medium with inhomogeneous EM properties to control the propagation of EM fields within the waveguide. Instead, the propagation of EM fields within the geodesic waveguide is governed by its geometry. Conversely, the disadvantages of the geodesic waveguide as an analog EM model of gravity comprise the following two aspects: firstly, losses may become significant upon considering finite-conductivity conductors at high frequency. Secondly, the experimental sampling of the electric field on the mid-surface requires a measurement system such as that used by Cheng \cite{cheng2010omnidirectional} in his omnidirectional absorber.

\section{Conclusions}
\label{sec:conclusions}

In this study, we have demonstrated that a geodesic waveguide, formed by a pair of axially symmetric curved parallel conducting surfaces, can be used as an analog EM model of a static and spherically symmetric gravitational field, such as those represented by a Schwarzschild BH and a Morris-Thorne WH. Our findings confirm that a one-dimensional Gaussian beam propagates within the geodesic waveguide following effective null-geodesics on its mid-surface, which, in turn, correlate with general relativistic solutions. These trajectories are equivalent to those which would have taken place in the flat waveguide. At the same time, as the flat waveguide is a simplification of an analog EM model of gravity based on a radially varying refractive medium, which, in turn, stems out from the space-time metric of a static and spherically symmetric gravitational field, the aforementioned trajectories are also equivalent to those taking place on the invariant plane of a static and spherically symmetric gravitational field. Moreover, the mean relative error between the trajectories followed by the Gaussian beam in the geodesic waveguide and the theoretical null-geodesics computed in the flat waveguide framework remains below 4\% in the cases where they can be compared. This result implies the applicability of the geodesic waveguide as an analog EM model of gravity.

In our simulations, some wave effects, like beam wavefront distortion and even beam splitting, have been observed. These phenomena are particularly strong as the beam propagates next to the analog photon sphere. However, in these cases, it was impossible to directly compare the path followed by the beam on the mid-surface and the solutions from the ray equation due to the distortions of the field distribution. Despite that, these observations suggest that the geodesic waveguide can be used not only with Gaussian beams but also with other field sources, like plane or cylindrical waves.

The limitations and advantages of our approach, alongside its comparison with the current literature, have already been clearly stated in Sec. \ref{sec:SoA}. Perhaps the most practical advantages are the fulfillment of space-time consistency conditions and the availability of fabrication possibilities and techniques in the microwave frequency range. The significance of this study is that the simplicity of the working principle of the geodesic waveguide makes either its physical or simulated application immediately available to researchers in other disciplines. 

\begin{acknowledgments}
The authors thank Millimeter wave Array at Room Temperature for Instruments in LEO Altitude Radio Astronomy (MARTINLARA) Project No. P2018/NMT-4333. V.D.F. acknowledges the Istituto Nazionale di Fisica Nucleare, Sezione di Napoli, iniziative specifiche TEONGRAV, and Gruppo Nazionale di Fisica Matematica of Istituto Nazionale di Alta Matematica. V.D.F. thanks the University Carlos III of Madrid for the hospitality and support in carrying out this work. The authors are grateful to the anonymous referee for the insightful comments aimed at improving the impact of the paper.
\end{acknowledgments}

\appendix
\section{The isotropic coordinates}\label{sec:isotropic_coordinates}

The interval of static and spherically symmetric gravitational fields is commonly expressed in the following form (standard spherical frame) \cite{weinberg1972gravitation}
\begin{equation} \label{eq:metric_GSIM_space_times}
\dd s^2 = g_{\rm tt}(r)c^2\dd t^2+g_{\rm rr}(r)\dd r^2+r^2\dd\Omega^2,
\end{equation}
where $\dd\Omega^2=\dd\theta^2+\sin^2 \theta \dd \phi^2$ is the line element of the unit sphere, and $\text{diag}[g_{\rm tt}(r),g_{\rm rr}(r),r^2,r^2 \sin^2 \theta ]$ is the metric tensor. 
Employing the following coordinate transformation \cite{falcon2022propagation,weinberg1972gravitation}:
\begin{equation}
\rho (r) = C_0  \exp\left[\int\limits_{r_{\rm ref}}^{r} \frac{\sqrt{g_{\rm rr}(r')}}{r'} dr'\right],
\label{eq:conformalTransformationstdtoiso}
\end{equation}
the above metric can be expressed in the so-called isotropic form as shown in Eq.~\eqref{eq:isotropic_metric_tensor}. For the Schwarzschild BH \cite{carroll2019spacetime,luminet1979image} ($r_{\rm ref}=r_{\rm s}$) with $r_{\rm s}$ being the Schwarzschild radius in standard coordinates, we obtain
\begin{equation}
\rho = \frac{1}{2} \left(r-\dfrac{r_{\rm s}}{2}+\sqrt{r^2-r \cdot r_{\rm s}}\right);
\end{equation}
while for the Morris-Thorne WH \cite{morris1988wormholes} ($r_{\rm ref}=b_{\rm o}$) with $b_{\rm o}$ being the WH mouth size in the standard frame, we obtain
\begin{equation}
\rho = \frac{1}{2} \left(r+\sqrt{r^2-b_{\rm o}^2}\right).
\end{equation}
It is worth mentioning that the constant $C_0$ must be retrieved from the asymptotic flatness boundary condition: $\lim_{\rho \to \infty} J(\rho) = 1$, where $\rho \to \infty$ also entails $r \to \infty$.

\section{The photon sphere}\label{sec:photon_sphere}
The ray equation \eqref{eq:ray_equation} can be reorganized as follows
\begin{equation}
\frac{1}{\rho ^2  n^2(\rho) }+\frac{1}{n^2(\rho) \rho ^4}\left( \frac{\text{d} \rho}{\text{d} \phi}\right)^2 = \frac{1}{L_{\phi}^2},\label{eq:geodesic_equivalent_problem}
\end{equation}
which resembles the law of energy conservation in kinematics. In particular, we recognize the effective potential $V_{\rm eff}$ and kinetic energies $K_{\rm eff}$ as follows:
\begin{subequations}
\begin{align}
V_{\rm eff} &= \frac{1}{\rho ^2  n^2(\rho) },\label{eq:effective_potential}\\
K_{\rm eff} &= \frac{1}{n^2(\rho) \rho ^4}\left( \frac{\text{d} \rho}{\text{d} \phi}\right)^2. 
\end{align}
\end{subequations}
The photon sphere is defined by the radial distance at which $\dfrac{\dd V_{\rm eff}}{\dd \rho} = 0$. Thus, for a Schwarzschild BH (cf. Eq.~\eqref{eq:refraction_index_schwarzschild}) we have
\begin{equation}
\rho_{\rm pho} = \left(2+\sqrt{3} \right) r_{\rm si};
\end{equation}
and for the Morris-Thorne WH (cf. Eq.~\eqref{eq:ellisRefractionIndex}) we obtain
\begin{equation}
\rho_{\rm pho} = b_{\rm oi},
\end{equation}

with $r_{\rm si} = r_{\rm s}/4$ and $b_{\rm oi} = b_{\rm o}/2$ being the analog event horizon and WH mouth, respectively.

\section{The pullback transformation}\label{appendix_C}
In this appendix, we present a formal demonstration of the pull-back map used in Sec.~\ref{sec:EM_analog} to map the constitutive matrices of the gravitational domain onto the constitutive tensors of the analog domain. Although the development presented here is utterly based on that of Thompson \cite{thompson2011completely,thompson2010dielectric}, there are two fundamental differences between our approaches: firstly, we do not use the Hodge dual of the excitation tensor. Secondly, a contravariant constitutive tensor is used instead of a mixed one.

Let us consider the gravitational domain. As mentioned, it consists of  a manifold $\mathcal{M}$ endowed with a metric $g_{\mu \nu}$. The Cartesian coordinates $x^\mu$ on $\mathcal{M}$ are employed to indicate the space-time points. Then, the EM configuration of the gravitational domain is given by tensors $F_{\mu \nu}$, $f^{\mu \nu}$, $J^{\mu}_{\text{den}}$, and $\chi^{\mu \nu  \alpha \beta}$. The matrix representation of the four first tensors is
\begin{subequations}
\begin{align} 
F_{\mu \nu}
&= \frac{1}{c}\begin{pmatrix}
0 & \mathcal{E}_{\rm x} & \mathcal{E}_{\rm y} & \mathcal{E}_{\rm z}  \\
-\mathcal{E}_{\rm x} & 0 & -c\mathcal{B}_{\rm z} & c\mathcal{B}_{\rm y}  \\
-\mathcal{E}_{\rm y} & c\mathcal{B}_{\rm z} & 0 & -c\mathcal{B}_{\rm x}  \\
-\mathcal{E}_{\rm z} & -c\mathcal{B}_{\rm y} & c\mathcal{B}_{\rm x} & 0  \\
\end{pmatrix},
\label{eq:electromagneticfieldstrengthtensorB}\\
f^{\mu \nu}
&= \frac{1}{c} \begin{pmatrix}
0 & -c\mathcal{D}^{\rm x} & -c\mathcal{D}^{\rm y} & -c\mathcal{D}^{\rm z}  \\
c\mathcal{D}^{\rm x} & 0 & -\mathcal{H}^{\rm z} & \mathcal{H}^{\rm y}  \\
c\mathcal{D}^{\rm y} & \mathcal{H}^{\rm z} & 0 & -\mathcal{H}^{\rm x}  \\
c\mathcal{D}^{\rm z}& -\mathcal{H}^{\rm y} & \mathcal{H}^{\rm x} & 0  \\
\end{pmatrix},
\label{eq:electromagneticfieldexcitationtensorB}\\
\mathcal{J}^{\mu}
&= \left(c \rho, \mathcal{J}^{\rm x}, \mathcal{J}^{\rm y}, \mathcal{J}^{\rm z}  \right),
\label{eq:electromagneticfieldexcitationtensorB}\\
{J}^{\mu}_{\rm den}
&= \sqrt{-g}\mathcal{J}^{\mu}.
\label{eq:electromagneticfieldexcitationtensorB}
\end{align}
\end{subequations}
Where $\mathcal{J}^{\mu}$ is the four-current vector. The last piece of the EM configuration is the constitutive tensor $\chi^{\mu \nu  \alpha \beta}$. Next, let us consider the analog domain which consists of another manifold $\mathcal{\hat{M}}$, which is endowed with a Minkowski metric $\eta _{\mu \nu}$. The coordinates $\hat{x}^\mu \in \mathcal{\hat{M}}$ are employed to label the space-time points, while its EM configuration is defined by $\hat{F}_{\mu \nu}$, $\hat{f}^{\mu \nu}$, $\hat{J}^{\mu}_{\text{den}}$, and $\hat{\chi}^{\mu \nu \alpha \beta}$. 

Secondly, the elements of both domains are to be related by a transformation $\mathcal{T}: \mathcal{\hat{M}} \to \mathcal{M}$ shown in Eq.~\eqref{eq:transformation_T}. Its pull-back, see Eq.~\eqref{eq:transformation_T*},  $\mathcal{T}^*: \mathcal{{M}} \to \mathcal{\hat{M}}$ operates in the opposite sense \cite{thompson2011completely,thompson2010dielectric}.

The pull-back transformation $\mathcal{T}^*$ has associated the transformation matrix $\Lambda ^\mu _\nu  = \frac{\partial \hat{x}^{\mu}}{\partial x^{\nu}} $,  whose Jacobian is $\Delta_{\mathcal{T}^*} = \left| \frac{\partial \hat{x}^{\mu}}{\partial x^{\nu}} \right|$. Likewise, the determinant of the transformation matrix of $\mathcal{T}$ is $\Delta_{\mathcal{T}} = \Delta^{-1}$. \\

We proceed now to derive the explicit form of the transformation relationships of the EM configurations in both domains. The action of the aforementioned trasnformation on the tensors of the EM configuration follows the rules of tensor transformation relationships, for example
\begin{subequations}
\begin{equation}
\hat{F}_{\alpha \beta} = \mathcal{T}^* ({F}^{\mu \nu}) = \frac{\partial x^\mu}{\partial \hat{x}^\alpha} \frac{\partial x^\nu}{\partial \hat{x}^\beta}  {F}_{\mu \nu}, \label{eq:transformation_rule_Fstrength_munu}
\end{equation}
\begin{equation}
\hat{f}^{\alpha \beta} = \mathcal{T}^* ({f}^{\mu \nu}) = \Delta_{\mathcal{T}^*} ^{-1}\frac{\partial \hat{x}^\alpha} {\partial {x}^\mu} \frac{\partial \hat{x}^\beta}{\partial {x}^\nu} {f}^{\mu \nu}.\label{eq:transformation_rule_fexcitation_munu}
\end{equation}
\begin{equation}
\hat{J}^{\alpha} = \mathcal{T}^* ({J}^{\mu}_{\text{den}})  =\Delta_{\mathcal{T}^*} ^{-1} \frac{\partial \hat{x}^\alpha} {\partial x^\mu}  {J}^{\mu }.\label{eq:transformation_rule_fexcitation_munu}
\end{equation}
\label{eq_transformation_rule_pull_back}
\end{subequations}
So, in the analog domain, the excitation tensor at a space-time point $\hat{x}^{\eta}$ is given by

\begin{equation}
\hat{f}^{\alpha \beta} \Big\rvert_{\hat{x}^{\eta}} = \frac{1}{2}  \hat{\chi}^{ \alpha \beta \mu \nu} \Big\rvert_{\hat{x}^{\eta}} \hat{F}_{\mu \nu} \Big\rvert_{\hat{x}^{\eta}},\label{appendix_C_eq_01}
\end{equation}

\noindent Similarly, in the gravitational domain the excitation tensor at a space-time point $\mathcal{T}(\hat{x}^{\eta})$

\begin{equation}
{f}^{\gamma \tau } \Big\rvert_{\mathcal{T}(\hat{x}^{\eta})} = \frac{1}{2} {\chi}^{ \gamma \tau \kappa \lambda} \Big\rvert_{\mathcal{T}(\hat{x}^{\eta})} {F}_{\kappa \lambda} \Big\rvert_{\mathcal{T}(\hat{x}^{\eta})}.\label{appendix_C_eq_02}
\end{equation}

\noindent Then, provided Eqs.~\eqref{eq_transformation_rule_pull_back}, and considering  Eq.~\eqref{appendix_C_eq_02}, Eq.~\eqref{appendix_C_eq_01} can be thus rewritten as follows

\begin{equation}
\frac{1}{2}  \hat{\chi}^{ \alpha \beta \mu \nu} \Big\rvert_{\hat{x}^{\eta}} \mathcal{T}^{*}({F}_{\kappa \lambda }\rvert_{\mathcal{T}(\hat{x}^{\eta})}) = \atop \mathcal{T}^{*}\left( \frac{1}{2} {\chi}^{ \gamma \tau \sigma \xi} \Big\rvert_{\mathcal{T}(\hat{x}^{\eta})} {F}_{\sigma \xi} \Big\rvert_{\mathcal{T}(\hat{x}^{\eta})}\right).\label{appendix_C_eq_03}
\end{equation}
\noindent As both side of the above equation correspond to a tensor density of weigth -1, its action on a tensor density $\hat{V}_{\alpha \beta}$ of weight +1 results in a scalar $\alpha \in \mathbb{R}$. Thus

\begin{equation}
\frac{1}{2}  \hat{\chi}^{ \alpha \beta \mu \nu} \Big\rvert_{\hat{x}^{\eta}} \mathcal{T}^{*}({F}_{\kappa \lambda }\rvert_{\mathcal{T}(\hat{x}^{\eta})}) \hat{V}_{\alpha \beta} \Big\rvert_{\hat{x}^{\eta}} = \atop \mathcal{T}^{*}\left( \frac{1}{2} {\chi}^{ \gamma \tau \sigma \xi} \Big\rvert_{\mathcal{T}(\hat{x}^{\eta})} {F}_{\sigma \xi} \Big\rvert_{\mathcal{T}(\hat{x}^{\eta})}\right) \hat{V}_{\alpha \beta} \Big\rvert_{\hat{x}^{\eta}} .\label{appendix_C_eq_04}
\end{equation}

Expressing the right side of the above equation as calculated in the gravitational domain should led to the same scalar, therefore

\begin{equation}
\frac{1}{2}  \hat{\chi}^{ \alpha \beta \mu \nu} \Big\rvert_{\hat{x}^{\eta}} \mathcal{T}^{*}({F}_{\kappa \lambda }\rvert_{\mathcal{T}(\hat{x}^{\eta})}) V_{\alpha \beta} \Big\rvert_{\hat{x}^{\eta}} = \atop  \frac{1}{2} {\chi}^{ \gamma \tau \sigma \xi} \Big\rvert_{\mathcal{T}(\hat{x}^{\eta})} {F}_{\sigma \xi} \Big\rvert_{\mathcal{T}(\hat{x}^{\eta})} \mathcal{T}\left( V_{\alpha \beta} \Big\rvert_{\hat{x}^{\eta}} \right).\label{appendix_C_eq_05}
\end{equation}

Then

\begin{equation}
\frac{1}{2}  \hat{\chi}^{ \alpha \beta \mu \nu} \Big\rvert_{\hat{x}^{\eta}}  \frac{\partial x^{\kappa}}{\partial \hat{x} ^\mu}  \frac{\partial x^{\lambda}}{\partial \hat{x} ^\nu}  {F}_{\kappa \lambda }\rvert_{\mathcal{T}(\hat{x}^{\eta})} \hat{V}_{\alpha \beta} \Big\rvert_{\hat{x}^{\eta}} = \atop  \Delta_{\mathcal{T}}\frac{1}{2} {\chi}^{ \gamma \tau \kappa \lambda} \Big\rvert_{\mathcal{T}(\hat{x}^{\eta})} {F}_{\kappa \lambda} \Big\rvert_{\mathcal{T}(\hat{x}^{\eta})} \frac{\partial \hat{x}^{\alpha}}{\partial x ^\gamma} \frac{\partial \hat{x}^{\beta}}{\partial x ^\tau} \hat{V}_{\alpha \beta} \Big\rvert_{\hat{x}^{\eta}} .\label{appendix_C_eq_06}
\end{equation}

removing $ {F}_{\kappa \lambda }$ and $\hat{V}_{\alpha \nu}$
\begin{equation}
 \hat{\chi}^{ \alpha \beta \mu \nu} \Big\rvert_{\hat{x}^{\eta}}  \frac{\partial x^{\kappa}}{\partial \hat{x} ^\mu}  \frac{\partial x^{\lambda}}{\partial \hat{x} ^\nu}  = \Delta_{\mathcal{T}}{\chi}^{ \gamma \tau \kappa \lambda} \Big\rvert_{\mathcal{T}(\hat{x}^{\eta})} \frac{\partial \hat{x}^{\alpha}}{\partial x ^\gamma} \frac{\partial \hat{x}^{\beta}}{\partial x ^\tau}  .\label{appendix_C_eq_07}
\end{equation}

Reorganizing the terms, we conclude that the constitutive tensor is mapped as a tensor density of weight -1

\begin{equation}
\hat{{\chi}}^{ \alpha \beta \mu \nu} \Big\rvert_{\hat{x}^{\eta}} = \Delta_{\mathcal{T}^*} ^{-1} \frac{\partial \hat{x}^\alpha}{\partial x^{\gamma}} \frac{\partial \hat{x}^\beta}{\partial x^{\kappa}} \frac{\partial \hat{x}^\mu}{\partial x^{\lambda}} \frac{\partial \hat{x}^\nu}{\partial x^{\tau}}  {\chi}^{ \gamma \kappa \lambda \tau} \Big\rvert_{\mathcal{T}(\hat{x}^{\eta})}       .\label{appendix_C_eq_08}
\end{equation}

\bibliography{references.bib}

\end{document}